# Enterprise Data Asset Quality: A Management–Standard Conformity–Benefit Realization Framework and Formation Mechanisms

Yixuan Zhu[1†], Aian Wu[1†], Runhan Zhang[1‡], Xiyi Wang[1‡], Junfeng Yu[1‡], Aihua Han[1*]

**Abstract:** Motivated by the limited standardization of enterprise data asset quality evaluation and the unclear relationship between assessment outcomes and value realization, this study develops a three-dimensional framework comprising *Data Asset Management Capability*, *Data Quality Standard Conformity*, and *Data Asset Benefit Realization Capability*, based on grounded theory and LDA topic modeling. To examine the formation mechanisms of data asset quality, this study adopts a multi-method approach combining PLS-SEM, Necessary Condition Analysis (NCA), and fuzzy-set Qualitative Comparative Analysis (fsQCA), to capture net effects, capability thresholds, and configurational paths. The results show that significant positive relationships exist among the three dimensions, with *Data Asset Management Capability* exerting the strongest effect on *Data Quality Standard Conformity* and further promoting *Data Asset Benefit Realization Capability*, forming a chain mechanism of *management foundation–standard enhancement–value realization*. In addition, all three dimensions constitute critical necessary conditions for achieving high data asset quality, and multiple equivalent configurational paths reflecting different combinations of *Management*, *Standard*, and *Benefit* are identified, such as governance-oriented and benefit-driven mechanisms. This study integrates structural (PLS-SEM), necessary-condition (NCA), and configurational (fsQCA) analyses within a unified framework, providing a systematic approach to understanding data asset quality formation and offering practical insights for enterprise data governance and data factor market development.

[1] School of Statistics and Mathematics, Zhongnan University of Economics and Law, Wuhan 430073, China
[†] These authors contributed equally to this work and share first authorship.
[‡] These authors contributed equally to this work and share second authorship.
[*] Corresponding author: Aihua Han, Email: aihuahan@zuel.edu.cn

# 1. Introduction

Driven by the global wave of digital economic transformation, data has evolved from an operational support resource into a core strategic asset that drives economic growth and corporate innovation. Advances in artificial intelligence, the Internet of Things, and big-data analytics have shifted data from a by-product of business processes to a direct source of value creation. The rise of big data has also reshaped research paradigms in production and operations management, emphasizing data-driven process redesign and capability development [1]. This shift fundamentally reinforces the practical relevance of "data as an asset", making data a key foundation for corporate decision-making and business impact [2]. Major economies have elevated the development of data factor markets to a national strategic priority. In the United States, the Federal Data Strategy frames data as a strategic asset to be governed and utilized, and promotes open government data and cross-sector sharing. The European Union, through the Data Act and the Data Governance Act, establishes harmonized rules for data sharing and circulation and advances a unified European data space. In Asia, Singapore's Trusted Data Sharing Framework provides a common language and systematic approach for trust-based data collaboration and innovation in the private sector.

Enterprises, as the primary carriers of data assets, play a central role in promoting the development and value realization of data elements [3]. Data asset quality, as a key prerequisite for the effective realization of data value, not only involves technical attributes such as accuracy, consistency, completeness, and timeliness, but also reflects the extent to which data supports organizational strategies, improves operational efficiency, and ensures compliant operations [4]. Compared with traditional data quality assessment, which mainly focuses on intrinsic technical attributes and emphasizes data operability and technical reliability [5], data asset quality assessment places greater emphasis on the overall performance of data in the process through which it is treated as an asset and supports business operations. Accordingly, the focus of evaluation shifts from "whether data is usable" to "whether data can support value realization" [6,7].

Data asset quality assessment is therefore not only a fundamental component of data governance, but also an important prerequisite for realizing the value of data elements, and has become a key issue of concern in both academia and practice. Existing studies have examined this issue from multiple perspectives, including data quality, data governance, and data asset evaluation, and have accumulated insights in indicator identification, weighting construction, industry-specific assessment, and pricing analysis [8–10]. However, there remains substantial room for further development in terms of conceptual integration, mechanism explanation, and cross-context applicability.

Specifically, first, regarding the object of evaluation, although prior studies have explored data quality, data governance, and data asset evaluation separately, there is still a lack of a unified definition of "data asset quality" as a composite concept encompassing technical, governance, and value attributes. In particular, existing research has yet to effectively integrate perspectives centered on data attributes and value realization, and a unified evaluation framework that simultaneously ensures standardization, operability, and value orientation remains underdeveloped [5–7]. Second, regarding explanatory mechanisms, existing studies mainly focus on indicator construction, weighting allocation, and evaluation outcomes, providing useful methodological support for data asset assessment [8,10]. However, they lack a systematic explanation of how enterprise data asset quality is formed through the transmission from management capability and standard conformity to benefit realization. Third, regarding applicability, most existing studies are conducted within specific industries, standards, or application contexts. While offering strong contextual relevance, they remain limited in terms of cross-industry comparison, unified evaluation criteria, and framework transferability [11,12], making it difficult to simultaneously address both generalizability and contextual adaptability in enterprise data asset quality assessment. Overall, existing research tends to focus on individual quality attributes, specific governance processes, or particular application scenarios, and has yet to simultaneously address conceptual definition, formation mechanisms, and cross-context applicability within a unified framework.

Building on the above discussion, this study addresses the central question of how enterprise data asset quality can be defined, measured, and its formation mechanisms explained. To further elaborate this inquiry, three interrelated research questions are proposed. First, how can operational and comparable evaluation dimensions of data asset quality be derived from enterprise practices and standard texts? Second, does the formation of data asset quality follow an internal structural logic among these dimensions? Third, does data asset quality exhibit differentiated configurational paths across different industrial contexts? To answer these questions, this study draws on survey data from data-element–oriented enterprises in Hubei Province. It first employs grounded theory and LDA topic modeling to construct and cross-validate a three-dimensional evaluation framework comprising *Data Asset Management Capability*, *Data Quality Standard Conformity*, and *Data Asset Benefit Realization Capability* (hereafter referred to as *Management*, *Standard*, and *Benefit*). On this basis, an evaluation model is developed using a combined weighting approach integrating AHP, the entropy weighting method, and CRITIC. Given that a single method is insufficient to capture the complex causal relationships underlying data asset quality formation, this study adopts a multi-method complementary design. Specifically, PLS-SEM is used to examine structural relationships among *Management*, *Standard*, and *Benefit*; Necessary Condition Analysis (NCA) is applied to identify capability thresholds; and fuzzy-set Qualitative Comparative Analysis (fsQCA) is employed to uncover configurational paths, thereby analyzing the formation mechanisms from the perspectives of net effects, necessary conditions, and configurational patterns.

Compared with existing studies, this study makes three main contributions. First, from the perspective of data assets rather than general data quality, it develops a three-dimensional evaluation framework integrating management capability, standard conformity, and value realization, thereby advancing the structured conceptualization of data asset quality. Second, beyond evaluation, it provides a systematic explanation of how data asset quality evolves from management capability to standard conformity and further to benefit realization, addressing the limitation of prior research that emphasizes evaluation while neglecting mechanism explanation. Third, under a unified evaluation framework, it examines capability thresholds and configurational differences across industries, thereby extending the applicability of data asset quality research in cross-context comparison. The findings offer practical implications for enterprise data governance and provide a reference framework for advancing data factor market development.

## 2. Theoretical Foundations of Data Asset Evaluation

### 2.1 Conceptual Definition

Unlike the general concept of data, data assets place greater emphasis on ownership and value attributes. Data can be regarded as assets when they are clearly owned or controlled and are capable of generating sustained economic benefits, in line with the Conceptual Framework for Financial Reporting [13]. It is further argued that enterprises should treat data as an important category of assets and emphasize its management and evaluation [10]. From the perspective of emerging production factors, Yin et al. suggest that data pricing is influenced by supply–demand mechanisms while also exhibiting distinctive characteristics such as nonrivalry [14]. The Data Asset Management Practice White Paper (Version 5.0) specifies that data assets include structured, semi-structured, and unstructured data, characterized by intangibility, replicability, and shareability [15].

Based on the above discussion, this study defines data assets as data sets that are created, owned, and controlled by firms in the course of production and operations, and that generate economic, strategic, or operational value. In essence, data assets are not merely intangible resources but composite assets characterized by economic value, strategic significance, and monetization potential.

A clear distinction should be made among three related concepts: data quality, data asset evaluation, and data asset quality assessment. Traditional data quality research focuses on intrinsic data attributes and emphasizes whether dimensions such as accuracy, completeness, and consistency meet usage requirements [16]. Data asset evaluation, by contrast, focuses on value recognition, valuation, and pricing, highlighting the economic attributes of data and its role in resource allocation [7,14]. In comparison, data asset quality assessment emphasizes the overall performance of data in the processes of effective governance, standard conformity, and value realization, integrating technical, governance, and value attributes [4,10]. Therefore, data asset quality assessment is neither a simple extension of traditional data quality research nor equivalent to value-oriented data asset evaluation, but rather represents a critical analytical layer linking data quality attributes with the formation of asset value.

## 2.2 Theoretical Construction of Data Asset Quality Dimensions

From the existing literature, related discussions mainly concentrate on three aspects: data governance, data quality, and data value. The first emphasizes institutional arrangements, organizational responsibilities, and resource allocation; the second focuses on attribute requirements such as accuracy, completeness, consistency, and timeliness; and the third examines the role of data in improving operational efficiency, optimizing decision-making, and enhancing competitive advantage. Although prior research provides an important foundation for understanding enterprise data management and value creation, it is generally conducted from a single-dimensional perspective and lacks a unified analytical framework that simultaneously integrates governance foundations, quality standards, and value outcomes.

Based on this, this study conceptualizes enterprise data asset quality as a composite system characterized by the coupling of three dimensions—*Management*, *Standard*, and *Benefit*—and further explores the internal relationships among them. Specifically, *Management* refers to the firm's capability in organizing, institutionalizing, and allocating resources for data assets, highlighting its institutional foundation and practical operability; *Standard* captures the extent to which data conforms to requirements such as accuracy, completeness, consistency, and timeliness, emphasizing its evaluability and comparability; and *Benefit* reflects the actual outcomes of data assets in operational optimization, decision support, and value realization, highlighting its outcome orientation and application performance. The naming of these three dimensions does not directly adopt general expressions such as *governance–quality–value*, but is instead derived from recurring themes identified in enterprise interview materials and standard documents, thereby ensuring closer alignment with the empirical context, indicator construction, and subsequent measurement logic of this study.

### 2.2.1 Data Asset Management Capability (*Management*)

*Data Asset Management Capability* constitutes the institutional and process foundation of the enterprise data asset quality evaluation system. Its core lies in ensuring the standardization and consistency of data assets throughout the entire lifecycle—including collection, processing, storage, sharing, and application—through appropriate organizational structures, institutional arrangements, and resource allocation. Theoretically, this dimension is grounded in Data Governance Theory. This theory emphasizes the establishment of formal mechanisms to clarify data-related rights and responsibilities, facilitate cross-departmental coordination, and implement unified processes and standards, thereby ensuring data availability, integrity, security, and compliance [17]. Inadequate governance systems may lead to data silos and hinder cross-departmental data flows, ultimately weakening the realization of data asset value [18]. Furthermore, from the perspective of the Resource-Based View (RBV), *Data Asset Management Capability* can be regarded as a dynamic capability that integrates internal data resources with the external environment, thereby generating organizational advantages that are difficult to imitate [19].

In this study, Data Governance Theory is primarily employed to define the institutional boundaries of the management dimension, emphasizing how enterprises provide a foundational governance basis for data asset operations through organizational arrangements, responsibility allocation, and resource coordination. The RBV perspective further explains why such a governance foundation can be transformed into the organizational capability required for subsequent standard development and value realization.

#### 2.2.2 Data Quality Standard Conformity (*Standard*)

*Data Quality Standard Conformity* emphasizes the technical attributes and evaluability of data and serves as a prerequisite for cross-system and cross-functional data sharing and circulation. The data quality model proposed by Wang and Strong provides the theoretical foundation for this dimension, encompassing key attributes such as accuracy, consistency, completeness, timeliness, and accessibility [16]. These attributes constitute the fundamental criteria for determining whether data can effectively support business requirements. From a semiotic perspective, the standards dimension plays a mediating role in transforming raw data into interpretable and meaningful assets. It is argued that recognizability and structured norms are essential prerequisites for achieving shared interpretation and value transformation across different contexts [5]. Accordingly, a data quality standards system functions as a key mechanism for enabling the symbolization of data and establishing shared semantics.

In this study, data quality theory is primarily employed to define the attribute boundaries of the standards dimension, that is, whether data satisfies basic requirements such as accuracy, consistency, and completeness. The semiotic perspective further explains why these standards are not only technical in nature, but also determine whether data can be consistently understood and effectively utilized across different business and institutional contexts.

#### 2.2.3 Data Asset Benefit Realization Capability (*Benefit*)

*Data Asset Benefit Realization Capability* emphasizes the role of data assets in generating economic value, supporting strategic and operational decision-making, and enabling innovation within organizations. This dimension is theoretically grounded in value chain theory and data economics. Prior research suggests that data do not inherently generate value; rather, value realization depends on organizational capabilities and governance structures [20]. From the perspective of value chain theory, Porter argues that competitive advantage arises from the coordination and optimization of activities across the value chain [21]. In this context, data assets contribute to value creation by enhancing operational processes, improving customer interactions, and supporting strategic decision-making. From the perspective of data economics, Brynjolfsson and McElheran conceptualize data as data capital, highlighting its role in improving productivity, enabling product differentiation, and enhancing innovation efficiency [22]. Laney further argues that data, as a quantifiable asset, should be evaluated based on its contributions to revenue growth, cost reduction, risk mitigation, and innovation outcomes [7].

In this study, value chain theory is primarily used to explain how data assets are embedded within organizational processes and transformed into operational and competitive advantages. Data economics further defines the outcome boundaries of the benefit dimension, emphasizing that the value of data should be reflected in observable outcomes—such as revenue, cost, risk, and innovation—rather than remaining at the level of potential resources.

#### 2.2.4 Logical Relationships and the Closed-Loop Mechanism

Building on the above dimensions, this study delineates the key causal relationships among *Management*, *Standard*, and *Benefit* in accordance with the explanatory scope of different theoretical perspectives. Data Governance Theory primarily informs the relationship between *Management* and *Standard*, emphasizing that governance arrangements constitute the institutional foundation for the formation and enforcement of data standards. The Resource-Based View (RBV) underpins the relationship between *Management* and *Benefit*, highlighting that management capability, as an organizational capability, can directly facilitate value creation. Value Chain Theory and Data Economics mainly inform the relationship between *Standard* and *Benefit*, suggesting that high-quality and transferable data can only be converted into observable performance outcomes when embedded in business processes and effectively utilized. Accordingly, the model proposes a primary causal chain characterized as *Management* as foundation – *Standard* as enhancement – *Benefit* as realization.

*Data Asset Management Capability* provides the institutional foundation for the establishment and enforcement of data standards. Through governance arrangements—such as organizational structures, formal rules, and resource coordination—firms ensure data security, integrity, and availability across the data lifecycle [23]. As governance systems mature, firms are better able to implement standardized procedures, clarify responsibility boundaries, and allocate resources more effectively, thereby enhancing *Data Quality Standard Conformity* [10].

*$H_1$: Data Asset Management Capability positively affects Data Quality Standard Conformity.*

*Data Asset Management Capability* also contributes directly to benefit realization by improving data flows, reducing redundancy, and strengthening security and compliance [24]. From an RBV perspective, management capability can be understood as a dynamic capability [25] that integrates internal and external data resources, thereby enabling innovation and supporting value transformation [26,27].

*$H_2$: Data Asset Management Capability positively affects Data Asset Benefit Realization Capability.*

*Data Quality Standard Conformity* constitutes a key condition for cross-system, cross-functional, and cross-scenario data circulation [16]. When data exhibit higher levels of accuracy, consistency, and completeness, they become more transferable and usable in both internal sharing and external exchanges [28]. As a result, standardization contributes to improved operational efficiency, reduced coordination costs, and enhanced innovation performance [27].

*$H_3$: Data Quality Standard Conformity positively affects Data Asset Benefit Realization Capability.*

### 2.3 Theoretical Positioning of the Data Asset Quality Evaluation Framework

Compared with existing studies and standard frameworks, the data asset quality evaluation framework proposed in this study does not seek to replace existing models, but rather to integrate the perspectives of data governance, data quality, and data value within a unified analytical framework.

Specifically, international standards such as ISO 8000 and ISO/IEC 25012 establish data quality requirements and evaluation criteria primarily around attributes such as accuracy, completeness, and consistency, with a focus on intrinsic data quality and conformity. *The Data Management Body of Knowledge (DAMA-DMBOK)* emphasizes governance structures, role allocation, knowledge domains, and best practices, providing systematic guidance for organizational data management; however, it does not in itself constitute an evaluation framework oriented toward value realization [4,23]. The Data Management Capability Maturity Model (DCMM) evaluates organizational data management from a maturity perspective, offering guidance for diagnosing and improving data management capabilities, but its application is more oriented toward organizational capability assessment and provides relatively limited support for cross-context unified evaluation [29].

Overall, existing frameworks provide important foundations in terms of quality attribute definition, governance system construction, and capability maturity assessment. However, they tend to focus on a single dimension—whether normative, managerial, or diagnostic—and rarely incorporate governance foundations, quality standards, and value outcomes into a unified analytical chain. As a result, while prior studies offer important references, they still lack an integrated framework capable of simultaneously linking conceptual definition, evaluation dimensions, and underlying mechanisms of data asset quality formation.

Based on this, the framework developed in this study aligns the Management dimension with the institutional foundations of data governance, the Standard dimension with data quality attributes and normative requirements, and the Benefit dimension with the outcomes of data assets in operational optimization, decision support, and value realization. The primary contribution of this framework does not lie in introducing new dimensions, but in integrating governance, quality, and value within a unified analytical framework and explicating their internal relationships, that is, a relational logic in which Management underpins Standard conformity, which in turn conditions Benefit realization. Furthermore, by combining enterprise practice materials, standard document analysis, and empirical examination, this study provides a more systematic explanation of the formation mechanism of data asset quality and offers a relatively consistent analytical entry point for comparative analysis across different industry contexts. Overall, the value of this framework lies in enhancing the integrative capacity, operational applicability, and explanatory power of data asset quality research.

# 3. Development of the Data Asset Quality Assessment Framework

## 3.1 Data Collection, Coding, and Robustness Verification

Enterprise data asset quality assessment requires generalizable insights derived from concrete business contexts. Accordingly, this study adopts grounded theory as the primary methodological approach. Grounded theory is a qualitative approach that relies on systematic data collection and iterative analysis to inductively develop theoretical constructs from empirical materials, and it is well suited to exploratory settings where extant theory remains incomplete [30]. Its bottom-up logic mitigates premature theoretical imposition and anchors the emergent framework in real organizational contexts. This choice aligns with the objective of developing a practice-anchored assessment framework [31,32], and prior work in this domain has similarly used grounded theory to develop evaluation frameworks [33]. Following established grounded-theory procedures, this study conducted open, axial, and selective coding to refine concepts and categories and derive the theoretical framework [34].

### 3.1.1 Data Collection: Sampling Data-Factor Enterprises

This study follows the grounded-theory principle of theoretical sampling, selecting subsequent cases based on concepts that emerge during iterative analysis to maximize theoretical development. To ensure diversity and representativeness, the study selects 13 Data-Factor Enterprises and conducts field investigations with semi-structured, in-depth interviews focused on data asset management practices, assessment standards, benefit realization pathways, and implementation challenges. To mitigate individual-response bias and strengthen informational triangulation, four employees from each enterprise were interviewed, covering functions such as data strategy, technical R&D, business application, and compliance. Each interview lasted 90–150 minutes, yielding nearly 30 hours of interviews and approximately 140,000 words of transcripts, which provide the empirical basis for subsequent coding.

### 3.1.2 Coding Procedures Based on Grounded Theory

NVivo was used to support coding, following the three-stage grounded-theory procedure. In open coding, the transcripts were examined line by line and conceptualized. Key segments describing data asset management phenomena, issues, or viewpoints were labeled, resulting in 1,525 initial codes. Through constant comparison, refinement, and consolidation, semantically similar codes were aggregated into higher-order concepts. Building on these concepts, categorization was conducted to derive 12 initial categories (zero-order categories). Table 1 provides illustrative examples of the coding process.

Table 1. Illustrative Examples of Open Coding

| **Example Text Segment** | **Labeling** | **Conceptualization** | **Zero-Order Category** |
|---|---|---|---|
| "Communication barriers are minimal. Most issues arise only when coordinating across internal and external units, and these do not materially affect data-processing efficiency. Everyone generates data, and there are few obstacles between departments." | Limited interdepartmental communication issues | Data communication and flow | Organizational structure of data governance |
| "Some data is specific to individual departments, and differences often arise in how the two sides interpret or describe them. Their definitions and standards may differ, and issues emerge during data exchange and usage." | Existence of data-definition inconsistencies | | |
| "With a well-established 'after-sales service' mechanism, communication-related questions can be effectively transformed into suggestions for improving data services." | Communication supported by an 'after-sales' mechanism | | |
| "The company has simplified its internal issue-reporting procedures, with each step assigned to dedicated personnel for processing." | Clear and simplified communication procedures | | |

Building on open coding, axial coding aims to identify and establish the underlying logical relationships among categories. Focusing on the core phenomenon, the causal and structural relationships among the zero-order categories were examined. The twelve zero-order categories were consolidated into four first-order categories: *Data Asset Management Capability, Data Quality Standard Conformity, Data Asset Benefit Realization Capability*, and External Moderating Factors of Data Assets (Table 2). While the three dimensions draw on established theoretical frameworks for boundary delineation, they were iteratively refined through the coding process rather than mechanically imposed from prior theory. Initial concepts were generated through open coding and subsequently refined through axial coding, which grouped related categories and identified their functional distinctions. Concepts related to organizational mechanisms, institutional arrangements, and resource coordination were categorized under the Management dimension. Concepts associated with accuracy, consistency, completeness, timeliness, and normative requirements were assigned to the Standard dimension. Concepts referring to operational improvement, decision support, and value realization outcomes were grouped under the Benefit dimension. As a result, the three-dimensional framework is grounded in empirical evidence and exhibits clearly defined conceptual boundaries.

Table 2. Axial Coding Results

| Zero-Order Category | First-Order Category |
|---|---|
| Organizational Structure of Data Governance<br>Resource Allocation for Data Governance<br>Governance Mechanisms for Data Management | Data Asset Management Capability |
| Data Quality Standards<br>Data Technology Tools<br>Multi-Source Data Integration | Data Quality Standard Conformity |
| Business Value<br>Strategic Value<br>Innovation Value | Data Asset Benefit Realization Capability |
| Policies and Regulations | External Moderating Factors of Data Assets |
| Industry–University Collaboration<br>Market and Users | |

Selective coding integrates all categories around a core category and clarifies their interrelationships to construct a coherent theoretical framework. Data Asset Quality is identified as the core category, from which an evaluation framework is derived comprising three core dimensions—*Data Asset Management Capability, Data Quality Standard Conformity, and Data Asset Benefit Realization Capability*—supplemented by External Moderating Factors. The management dimension forms the foundation of data asset governance, encompassing organizational structure, resource allocation, and governance mechanisms that ensure operational consistency. The *standard* dimension reflects data evaluability and includes attributes such as accuracy, consistency, completeness, and accessibility, which determine data reliability within business processes. The benefits dimension captures the value-realization outcomes of data asset management, including operational efficiency, technological innovation, and strategic support. External Moderating Factors, such as policy environment, industry regulation, university–enterprise collaboration, and market feedback, continuously influence the three core dimensions, enhancing the framework's dynamic adaptability. The *Management* dimension provides the institutional basis for *Standard*; *Standard* constitutes the precondition for *Benefit*; and realized benefits validate the effectiveness of governance and standardization. External Moderating Factors dynamically shape the interactions among the three dimensions, ensuring applicability across diverse organizational and industry contexts.

#### 3.1.3 Theoretical Saturation and Coding Reliability

Theoretical saturation refers to the stage at which additional data no longer generate new properties, dimensions, or relationships, indicating that the theoretical structure has stabilized [34]. After completing the three coding stages using the first 75% of the interview transcripts, the core categories and their relationships were found to be stable. The remaining 25% of the transcripts were then analyzed as a validation set to assess saturation. No new concepts or categories emerged, and the existing framework adequately accounted for the additional data, indicating that theoretical saturation had been achieved.

Coding reliability was assessed to ensure objectivity and consistency. An additional qualitative researcher independently coded the raw transcripts. Cohen's Kappa was then calculated to evaluate inter-coder agreement. The resulting Kappa value of 0.90 indicates a high level of coding reliability.

### 3.2 Text-Mining Results Based on the LDA Topic Model

To validate the first-order framework derived from grounded theory and refine a second-order indicator system, an LDA topic model was applied to conduct unsupervised topic extraction from domestic and international industry standards related to data assets. This approach integrates qualitative insights with quantitative text mining to enhance methodological robustness.

Grounded theory and LDA exhibit two forms of methodological complementarity. First, they draw on complementary data sources: grounded theory relies on interview data that capture tacit organizational practices, whereas LDA analyzes industry standards that reflect explicit and codified sectoral norms. Comparing results across these sources enables cross-validation and strengthens the alignment between practice and industry standards [35,36]. Second, grounded theory follows an inductive logic in which theoretical categories emerge from empirical cases. LDA, by analyzing standardized texts, provides a form of deductive verification by examining whether inductively derived themes appear consistently across a broader corpus. Prior research indicates that integrating grounded theory with topic modeling is an effective approach for constructing and validating theoretical frameworks [37]. By extracting shared concepts and high-frequency themes from industry standards, the analysis identifies core dimensions of Data Asset Quality emphasized across governance contexts, providing substantive support for the evaluation framework.

#### 3.2.1 Model Selection and Theoretical Foundations of LDA

Latent Dirichlet Allocation (LDA), proposed by Blei et al. [38], is a probabilistic generative topic model based on the bag-of-words assumption that identifies latent thematic structures in unstructured text. Operating within a words–topics–documents structure and employing Dirichlet priors for topic generation, LDA has been widely applied in text analysis and indicator construction across domains [39,40]. Compared with LDA, Latent Semantic Analysis (LSA) performs dimensionality reduction but offers limited semantic interpretability. Correlated Topic Models (CTM) capture inter-topic dependencies but are sensitive to topic-number specification and may lead to semantic overgeneralization. Dynamic Topic Models (DTM) emphasize temporal evolution and are therefore less suitable for the static corpus examined in this study. Neural topic models such as the Embedded Topic Model (ETM) require substantial computational resources and rely heavily on word embeddings, reducing their methodological fit for the present analysis. Considering interpretability, stability, and methodological fit, LDA is selected as the topic-modeling approach.

#### 3.2.2 Data Sources and Text Preprocessing

The dataset consists of 83 industry-standard documents, including 21 Chinese and 62 English standards, covering domains such as Data Asset Management, information-system development, data quality control, and data privacy governance. All documents were converted from PDF to TXT format using Python and subjected to systematic preprocessing and noise removal. The core documents are listed in Table 3.

Table 3. Major Industry Standards Included in the Analysis

| Standard Name | Brief Description | Year of Release |
| --- | --- | --- |
| DCMM (Data Management Capability Maturity Model) | Provides a maturity assessment framework for enterprise data-management capabilities, helping organizations identify their developmental stage and improve data-management performance. | 2018 |
| Guidelines for Digital Asset Evaluation | Standardizes the evaluation of digital assets, including asset classification, valuation methods, management requirements, and risk-control mechanisms. | 2023 |
| Information Technology—Big Data—Data Asset Value Assessment | Specifies requirements for the management of big-data assets, including data-quality control, lifecycle management, and data-security practices. | 2021 |
| Information Technology Service—Data Asset—Management Requirements (GB/T 40685-2021) | Defines data-asset management requirements in IT service contexts, covering data storage, processing, sharing, and protection. | 2021 |

(Continuation)

| Standard Name | Brief Description | Year of Release |
| --- | --- | --- |
| Informationized Asset Management—Data Quality Management Requirements | Provides guidelines for data-quality control and evaluation in information-management systems, emphasizing standardization and quality-improvement methods. | 2024 |
| ISO 8000 Series | International standards covering data management, data quality, data governance, and data protection, including requirements for data quality, security, and auditing. | 2012–2024 |
| ISO 27001 Series | International standards for information security management systems (ISMS), addressing data security, privacy protection, access control, encryption, and breach prevention. | 2022–2024 |
| GDPR (General Data Protection Regulation) | EU regulation governing the processing, storage, transfer, and sharing of personal data, emphasizing data sovereignty, transparency, and data-subject rights. | 2016 |
| COBIT (Control Objectives for Information and Related Technologies) | A governance framework for information systems, outlining best practices for IT control, data-quality assessment, and IT-risk management. | 2018 |

For the Chinese corpus, jieba was used for word segmentation, and invalid tokens were removed using a stop-word list and single-character filtering rules. For the English corpus, a customized preprocessing pipeline was implemented, including lowercasing, removal of punctuation and digits, stop-word filtering, and exclusion of single-character tokens. Both corpora underwent three rounds of manual review, resulting in a standardized corpus suitable for LDA analysis. As LDA is an unsupervised learning method that does not require train–test partitioning, the full corpus was included in model estimation.

**3.2.3 Text Preprocessing, Tool Selection, and Parameter Configuration**

The LDA model was implemented using the gensim library. The library employs variational inference and is suitable for small- to medium-scale corpora. Alternative implementations, including sklearn and Mallet, were tested; considering computational efficiency and stability, gensim was selected.

In the LDA topic model, for $K$ topics and a document $d$, the probability of generating a word $w$ in document $d$ is given by:

$$P(w) = \prod_{n=1}^{N} \sum_{z=1}^{K} P(w_n|z;\phi)P(z|d;\theta) \tag{1}$$

where $\theta \sim \text{Dirichlet}(\alpha)$, $\phi \sim \text{Dirichlet}(\beta)$. Here, $P(w_n|z;\phi)$ denotes the probability of word $w_n$ under topic $z$, and $P(z|d;\theta)$ represents the topic distribution for document $d$.

Default Dirichlet priors were adopted for hyperparameters $\alpha$ and $\beta$. The number of training iterations was set to 100 to ensure convergence. Topic selection was guided by perplexity and coherence metrics. Candidate topic numbers ranging from 2 to 15 were evaluated, and metric curves were plotted for both Chinese and English corpora. The optimal number of topics was determined to be 3 for the Chinese standards and 4 for the English standards.

3.2.4 LDA Results and Indicator-System Refinement

The LDA coherence scores were 0.7182 for the Chinese corpus and 0.8033 for the English corpus, both exceeding the threshold of 0.7. Topic labeling and categorization were conducted manually based on term frequency, topic representativeness, and semantic clarity. The resulting themes were consistent with the first-order dimensions—*Data Asset Management Capability, Data Quality Standard Conformity, and Data Asset Benefit Realization Capability*—and further refined their subdimensions. The three-dimensional framework is not solely derived from interview-based inductive analysis, but is also consistently reflected in relatively stable topic clustering across both enterprise practice materials and standard documents. Grounded theory identifies dimensional boundaries from practical contexts, whereas the LDA model examines whether these boundaries exhibit consistency and interpretability in external normative texts, providing complementary empirical support for the three-dimensional structure.

The Chinese-topic structure closely matched the first-order categories, whereas the English results exhibited partial overlap. Across both corpora, high-frequency terms such as *accuracy, completeness, consistency, regulation,* and *accessibility* were identified. These patterns suggest semantic consistency across language-specific standards. The findings are consistent with the technical data-quality attributes defined in *ISO/IEC 25012* (Data Quality Model) and in *China's Data Quality Management Requirements for Informationized Asset Management* (2024).

Based on topic labeling and indicator abstraction, a set of evaluable second-order dimensions was derived. These dimensions correspond to key constructs such as strategy, governance, accuracy, consistency, business impact, and decision support. Together, they form a comprehensive indicator system aligned with the theoretical framework (Table 4).

Table 4. Second-Order Indicator Framework

| Primary Indicator | Secondary Indicator | LDA-Derived Keywords |
|---|---|---|
| Data Asset Management Capability | Data Strategy | *business, policy, development, market, objective, execution, direction* |
| | Data Governance | *governance, controller, oversight, supervisory, risk, control, compliance* |
| | Data Architecture | *design, systems, service, technical architecture, blockchain, class* |
| | Data Application Capability | *processing, data item, activity, implementation, software, service* |
| | Data Security & Privacy Protection | *security, protection, privacy, risk, access control* |
| | Data Quality Management | *quality, accuracy, completeness, measurement, standards* |
| | Data Standards & Specifications | *standards, criteria, specification, model, compliance* |
| Data Quality Standard Conformity | Accuracy | *accuracy, measurement, performance, validation* |
| | Consistency | *consistency, criteria, reliability, control* |
| | Completeness | *completeness, measurement, representation* |
| | Compliance | *compliance, regulation, legal, procedure* |
| | Timeliness | *timeliness, update, change, process* |
| | Accessibility | *access, system, user, procedure* |
| Data Asset Benefit Realization Capability | Business Impact | *business, commercial, value, economic, market* |
| | Decision Support Value | *decision, support, management, assessment* |
| | Innovation & Opportunity Creation | *innovation, opportunity, technology, trend, trust, contracts* |
| | Revenue Generation | *revenue, value, profit, cost, transaction* |
| | Cost Savings | *cost, savings, efficiency, reduction* |
| | Competitive Advantage | *competitive, advantage, market, differentiation, risk management* |
| | Risk Management & Control | *risk, control, security, governance, legal* |

### 3.3 Integrating a DCMM – Semiotics Perspective

Semiotics examines how meaning is constructed and transmitted through signs, emphasizing polysemy and contextual dependence. These properties provide analytical insight into how data assets are represented and utilized across industries and institutional settings. Polysemy implies that identical data may generate different referents across contexts, whereas contextual dependence requires alignment between data evaluation and specific institutional and industry environments. Data must undergo structuring, processing, and semantic encoding before raw information becomes value-embedded data assets [41]. Accordingly, Data Asset Quality depends not only on technical attributes but also on the capacity for effective symbolization and value realization within context-specific settings. A semiotic perspective addresses the persistent challenge of standard unification in enterprise Data Asset Quality evaluation. Within this framework, polysemy supports the development of shared indicator logic, whereas contextual dependence enables industry-specific adaptation without undermining generalizability. This balance enhances the applicability of the evaluation system.

To strengthen contextual interpretability, Morris's triadic structure—semantics, syntactics, and pragmatics—is incorporated into the third-order framework. The semantic dimension focuses on meaning construction and corresponds to the third-order indicators capturing critical aspects of Data Asset Quality. The syntactic dimension addresses structural organization and is reflected in the evaluation tool's three components: preliminary screening and cognitive assessment, quantitative scoring of core dimensions, and enterprise background information. The pragmatic dimension emphasizes contextual suitability within specific industries and application scenarios, enabling context-sensitive evaluation based on operational realities and assessing how indicators function in practice.

At the first and second order levels, widely used international data-quality standards and management frameworks are integrated to enhance structural comparability. Building on this structure, the eight capability domains and five maturity levels of *DCMM* are incorporated to refine the third-order specification of the management dimension. It provides a systematic structure for assessing current data management capability and defining improvement paths. It covers eight domains, including organization, strategy, standards, security, and quality, and specifies five maturity levels [42]. To examine the consistency between the categories derived from grounded theory and the DCMM framework, this study maps the categories generated from interview coding onto the eight capability domains of DCMM on a one-to-one basis, with independent verification conducted by two researchers. A category is considered a valid match if it demonstrates stable correspondence with the respective capability domain in terms of core content, managerial object, and implementation stage. The degree of structural overlap is then measured as the proportion of validly matched categories relative to the total number of categories compared. The results show that this proportion exceeds 80%, suggesting that the management dimension not only reflects enterprise practices to a considerable extent but is also broadly consistent with the national standard framework. Ultimately, drawing on the framework proposed by Dong and Zhang concerning the value logic and attribute characteristics of data assets [43], this study develops an enterprise data asset quality evaluation framework that integrates the dual perspectives of DCMM and semiotics. At the theoretical level, the framework provides a unified analytical perspective for understanding data asset quality across different contexts. At the methodological level, it reinforces the hierarchical structure and operational applicability of the evaluation system. At the practical level, it offers a structured basis for standardized indicator construction and context-specific adaptation.

Overall, the enterprise Data Asset Quality evaluation framework follows a layered logic. The first-order structure—*Data Asset Management Capability, Data Quality Standard Conformity, and Data Asset Benefit Realization Capability*—derives from the integration of data governance theory, the resource-based view, and value chain theory. The second-order framework is derived through triangulation between grounded theory and LDA modeling, translating the abstract first-order dimensions into actionable domains, such as strategy, governance, accuracy, and business impact—based on both interview data and industry standards. The third-order framework operationalizes these domains by decomposing each second-order indicator into measurable practice-level items aligned with *DCMM* capability domains. The incorporation of semiotic principles ensures contextual grounding alongside cross-industry applicability. Together, these layers form an integrated an evaluation chain structured as: theoretical dimensions (Level 1), practical levels (Level 2), and specific measurement items (Level 3). This integrated design enables comprehensive coverage of the structural, semantic, and pragmatic dimensions of enterprise data assets, and establishes a foundation for subsequent cross-industry comparison and contextual adaptation at both the framework and indicator levels.

### 3.4 Weighting Methodology

Building on the first- and second-order indicators, the third-order framework is refined to enable empirical measurement of Data Asset Quality. A three-tier weighting mechanism integrating the Analytic Hierarchy Process (AHP), the Entropy Weight Method, and the CRITIC method is adopted to allocate weights across indicator levels. At the first-order level, given the conceptual heterogeneity among *Data Asset Management Capability, Data Quality Standard Conformity, and Data Asset Benefit Realization Capability*, AHP is employed based on pairwise evaluations provided by 15 experts. At the second-order level, where inter-indicator correlation is relatively low, the Entropy Weight Method is applied to compute objective weights, assigning greater weight to indicators with higher information entropy. At the third-order level, where indicator scores exhibit stronger intercorrelations, the CRITIC method (Criteria Importance Through Intercriteria Correlation) is employed. This method accounts for both indicator variance and inter-indicator conflict when determining weights. This combined weighting structure follows prior applications in the data asset domain [44,45], integrating expert judgment with data-driven weighting procedures. Within the AHP procedure, the judgment matrix for the three first-order dimensions is presented below.

$$\begin{bmatrix} 1 & \frac{5}{3} & \frac{5}{2} \\ \frac{3}{5} & 1 & \frac{3}{2} \\ \frac{2}{5} & \frac{2}{3} & 1 \end{bmatrix} \tag{2}$$

Based on the integrated AHP–Entropy–CRITIC weighting process, the resulting Data Asset Quality evaluation framework and corresponding weight distribution are reported in Table 5 (see Appendix for detailed indicator definitions).

Table 5. Enterprise Data Asset Quality Evaluation Framework

| Primary Indicator | Secondary Indicator | Weight | Tertiary Indicator | Weight |
|---|---|---|---|---|
| Data Asset Management Capability (0.5) | Data Strategy | 0.092 | Data Strategy Planning System | 0.415 |
| | | | Data Strategy Implementation System | 0.328 |
| | | | Data Strategy Evaluation System | 0.257 |
| | Data Governance | 0.101 | Data Governance Organizational System | 0.248 |
| | | | Data Governance Policy System | 0.250 |
| | | | Data Governance Communication Mechanism | 0.502 |
| | Data Architecture | 0.118 | Data Modeling System | 0.139 |
| | | | Data Distribution System | 0.187 |
| | | | Data Integration and Sharing Function System | 0.343 |
| | | | Metadata Management System | 0.330 |

(Continuation)

| Primary Indicator | Secondary Indicator | Weight | Tertiary Indicator | Weight |
|---|---|---|---|---|
| Data Asset Management Capability (0.5) | Data Application Capability | 0.139 | Data Analytics System | 0.527 |
| | | | Data Open-Sharing System | 0.231 |
| | | | Data Service System | 0.242 |
| | Data Security & Privacy Protection | 0.107 | Data Security Strategy System | 0.332 |
| | | | Data Security Management System | 0.222 |
| | | | Data Security Audit System | 0.446 |
| | Data Quality Management | 0.126 | Data Quality Objectives | 0.259 |
| | | | Data Quality Inspection | 0.317 |
| | | | Data Quality Analysis | 0.232 |
| | | | Data Quality Improvement | 0.191 |
| | Data Standards & Specifications | 0.113 | Business Glossary System | 0.167 |
| | | | Reference Data and Master Data System | 0.432 |
| | | | Data Meta-System | 0.250 |
| | | | Indicator Data System | 0.152 |
| | Data Lifecycle Management | 0.118 | Data Requirements Definition | 0.366 |
| | | | Data Design and Development System | 0.201 |
| | | | Data Operations and Maintenance System | 0.152 |
| | | | Data Retirement Management System | 0.282 |
| | Data Integrated Environment | 0.086 | Organizational Emphasis on Data Assets | 0.256 |
| | | | Appropriate Staffing for Data Asset Management | 0.131 |
| | | | Importance of Data Assets to Business Processes | 0.228 |
| | | | Maturity of Data Asset Management Policies | 0.138 |
| | | | Strength of Data Asset Management Compliance | 0.091 |
| | | | Market Competitiveness Derived from Data Assets | 0.154 |
| Data Quality Standard Conformity (0.3) | Accuracy | 0.164 | Data Fully Validated and Verified | 0.163 |
| | | | Absence of Obvious Data Errors | 0.218 |
| | | | Regular Accuracy Checks | 0.319 |
| | | | No Human Error During Data Collection | 0.175 |
| | | | Accuracy Meets Business Requirements | 0.125 |
| | Consistency | 0.146 | Data Consistency Across Systems | 0.166 |
| | | | Data Consistency Validation | 0.193 |
| | | | Duplicates Eliminated During Integration | 0.151 |
| | | | Real-Time and Accurate Data Synchronization | 0.140 |
| | | | Comprehensive Consistency Management Mechanism | 0.349 |
| | Completeness | 0.173 | Data Complete with No Missing Values | 0.143 |
| | | | All Required Information Recorded During Collection | 0.325 |
| | | | All Key Attributes Included | 0.190 |
| | | | Completeness Checks Conducted | 0.117 |
| | | | Completeness Meets Business Requirements | 0.224 |

(Continuation)

| Primary Indicator | Secondary Indicator | Weight | Tertiary Indicator | Weight |
|---|---|---|---|---|
| Data Quality Standard Conformity (0.3) | Compliance | 0.169 | Data Comply with Industry Standards | 0.171 |
| | | | Standardized Data Processing | 0.318 |
| | | | Unified Data Formats | 0.153 |
| | | | Adherence to Internal Data Management Rules | 0.136 |
| | | | Standardization Level Meets Business Needs | 0.222 |
| | Timeliness | 0.159 | Real-Time Data Updates | 0.154 |
| | | | Timely Data Collection and Processing | 0.363 |
| | | | Timeliness Meets Business Requirements | 0.216 |
| | | | Regular Timeliness Evaluation | 0.121 |
| | | | Data Reflect Business Changes in a Timely Manner | 0.145 |
| | Accessibility | 0.190 | Users Can Access Needed Data Anytime | 0.193 |
| | | | Ease of Data Discovery through Simple Queries | 0.232 |
| | | | Reasonable Access Permissions | 0.179 |
| | | | Adequate Documentation Supporting Understanding and Use | 0.230 |
| | | | No Technical Barriers in Accessing Data | 0.165 |
| Data Asset Benefit Realization Capability (0.2) | Business Impact | 0.193 | Improve Product/Service Quality and Reliability | 0.148 |
| | | | Increase Sales or Customer Satisfaction | 0.285 |
| | | | Expand Market Share or Business Scope | 0.261 |
| | | | Enhance Operational Efficiency or Delivery Speed | 0.122 |
| | | | Improve Understanding of Customer Needs or Market Trends | 0.184 |
| | Decision Support Value | 0.192 | Support Strategic Decision-Making | 0.212 |
| | | | Enable More Informed Competitive Decisions | 0.145 |
| | | | Improve Resource Allocation or Investment Decisions | 0.212 |
| | | | Optimize Supply Chain or Inventory Management | 0.250 |
| | | | Provide Critical Support During Crisis Management | 0.181 |
| | Innovation & Opportunity Creation | 0.098 | Facilitate New Product/Service Development | 0.249 |
| | | | Identify New Market Opportunities or Business Models | 0.184 |
| | | | Enable Innovations in Marketing or CRM | 0.157 |
| | | | Create Opportunities for Partnerships and Cross-Department Collaboration | 0.260 |
| | | | Support Digital Transformation or Strategic Innovation | 0.150 |
| | Revenue Generation | 0.105 | Direct increase in corporate revenue or sales | 0.239 |
| | | | Optimization of pricing strategies or marketing effectiveness | 0.179 |
| | | | Enhancement of customer lifetime value or repurchase rate | 0.261 |
| | | | Contribution to cross-selling or upselling activities | 0.133 |
| | | | Development of new revenue streams or value-added services | 0.187 |

(Continuation)

| Primary Indicator | Secondary Indicator | Weight | Tertiary Indicator | Weight |
|---|---|---|---|---|
| Benefit Realization Capability (0.2) | Cost Reduction | 0.107 | Reduction of operational or production costs | 0.141 |
| | | | Optimization of procurement management or supply chain costs | 0.237 |
| | | | Reduction of human resource management or labor costs | 0.142 |
| | | | Reduction of technology or IT infrastructure maintenance costs | 0.317 |
| | | | Cost savings in energy utilization or environmental management | 0.163 |
| | Competitive Advantage | 0.108 | Enhancement of brand recognition or reputation in the market | 0.215 |
| | | | Differentiation from competitors in product or service quality | 0.183 |
| | | | Enhance Customer Experience or Relationship Management | 0.184 |
| | | | Strengthen Market Positioning or Brand Differentiation | 0.121 |
| | | | Support Geographic Expansion or Market Penetration | 0.296 |
| | Risk Management | 0.100 | Identify and Predict Market or Industry Risks | 0.159 |
| | | | Improve Compliance or Legal Risk Management | 0.163 |
| | | | Manage Supply Chain and Supplier Risks | 0.284 |
| | | | Enhance Information Security and Data Privacy Protection | 0.195 |
| | | | Support Crisis Management and Business Continuity Planning | 0.199 |
| | Stakeholder Benefit | 0.096 | Improve Customer Product/Service Experience | 0.098 |
| | | | Improve Employee Efficiency and Satisfaction | 0.448 |
| | | | Increase Profitability for Shareholders | 0.123 |
| | | | Enhance Supplier Performance and Cooperation | 0.206 |
| | | | Strengthen Corporate Reputation and Social Responsibility | 0.125 |

# 4. Data Asset Quality Assessment Results

To examine the applicability of the evaluation framework, a field survey was conducted among Data-Factor Enterprises in Hubei Province, with 1,500 questionnaires distributed. The survey design emphasized representativeness and data reliability. Guided by the *"Data Elements ×" Three-Year Action Plan* (2024–2026), the sample was classified into five sectors: healthcare, cultural tourism, technology innovation, commercial circulation, and urban governance. This classification enables cross-sector comparison of assessment outcomes and provides a structural basis for interpreting score differentials. The composite assessment results indicate mean scores of 3.51 for *Data Asset Management Capability*, 3.81 for *Data Quality Standard Conformity*, 3.66 for *Data Asset Benefit Realization Capability*, and 3.82 for sample-weighted overall Data Asset Quality. These results suggest that enterprise Data Asset Quality in Hubei Province is at an intermediate level of development. Scores for standard conformity exceed those for management capability and benefit realization, indicating room for improvement in translating governance capacity into value outcomes.

Specifically, in the healthcare sector, clinical and health data are used to optimize diagnosis and treatment processes, support pharmaceutical R&D, and improve public health management. The cultural tourism sector leverages data elements to develop digital cultural products, enhance visitor experience, and support cultural heritage preservation. The technology innovation sector emphasizes the integration and reuse of scientific and R&D data to facilitate research transformation and knowledge commercialization. The commercial circulation sector applies data analytics to optimize supply chains and improve demand forecasting. The urban governance sector integrates urban operational data to enhance public services such as transportation, emergency response, and environmental management. The sectoral assessment results for enterprise Data Asset Quality are presented in Figure 1.

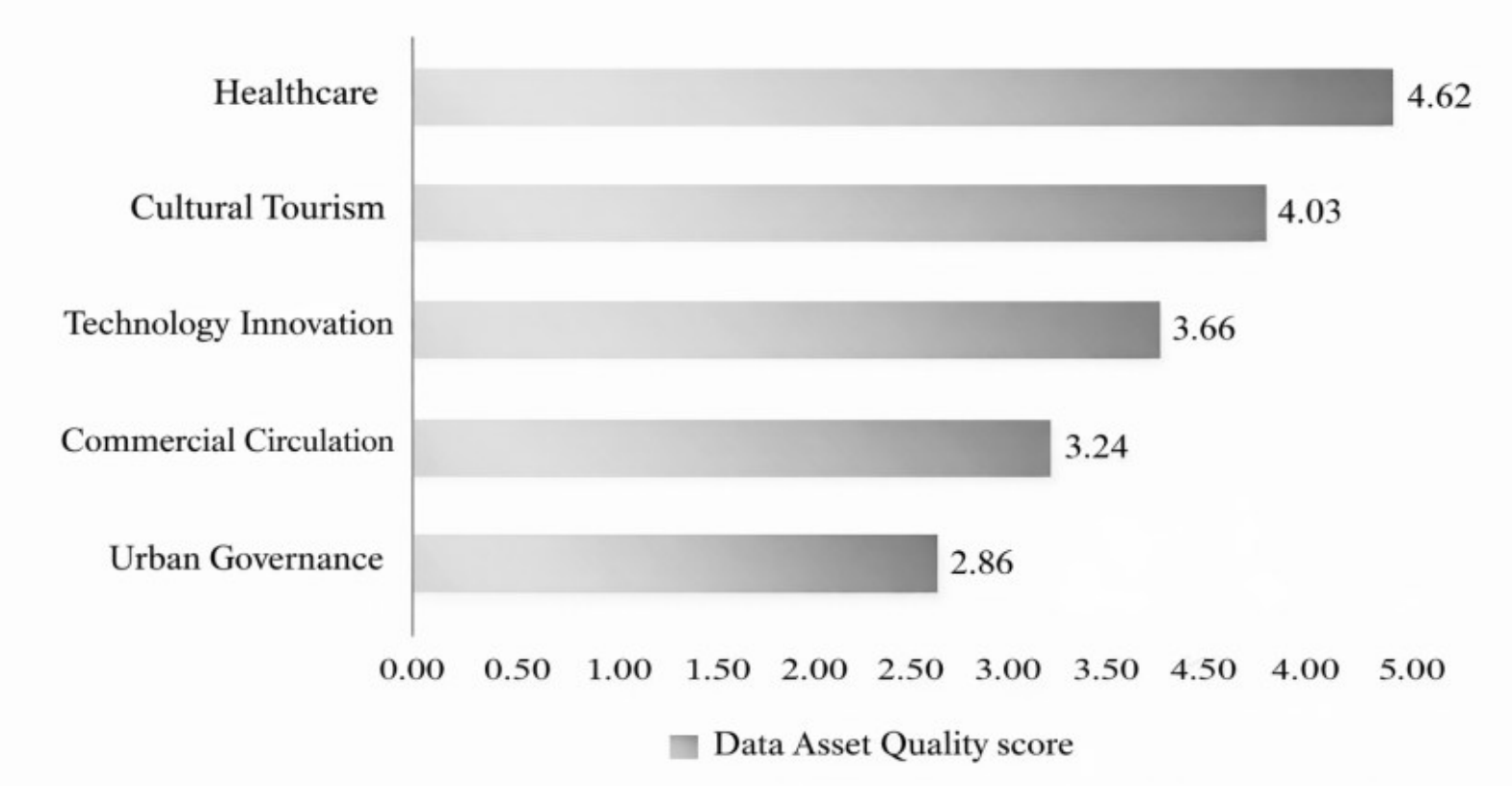

Figure 1. Data Asset Quality Scores Across the Five “Data Elements ×” Sectors

Figure 1 indicates an average Data Asset Quality score of 3.68 across the five sectors, with a median of 3.66. Given that the sampled firms are actively engaged in data-asset practices, the overall level remains moderate. The range of 1.76 suggests substantial variation across sectors. The Healthcare sector records the highest score at 4.62. The Commercial Circulation sector scores 3.24, ranking fourth, while the Urban Governance sector reports the lowest score at 2.86.

The Healthcare sector is characterized by highly standardized and regulated data environments, with stringent compliance requirements and relatively mature technical infrastructures, which are associated with higher quality scores. The Commercial Circulation sector involves multi-entity data environments with fragmented ownership structures and limited standardization in data collection, contributing to siloed and heterogeneous data formats. Reliance on traditional operational models and limited automation in data acquisition further constrain data quality, while resource-intensive processing of unstructured data reduces timeliness and practical usability. In addition, limited inter-firm data sharing within supply chains and cost-oriented operational priorities may weaken incentives for systematic data governance investment. The Urban Governance sector includes a substantial proportion of state-owned enterprises, where digital transformation processes may progress at varying speeds. Organizational hierarchies, risk-control priorities, and performance systems that emphasize operational outputs over data management may contribute to lower incentives for cross-departmental data integration and quality improvement.

While most findings are consistent with prior expectations, an unexpected pattern emerges: the cultural tourism sector ranks second with a score of 4.03, exceeding the Technology Innovation sector (3.66). This difference may reflect contrasting data utilization logics across sectors. In the Technology Innovation domain, particularly in large-model development, firms often emphasize data scale and diversity. Once datasets reach substantial volume, incremental improvements in individual data quality may yield comparatively limited returns. In addition, large-model training frequently relies on extensive web-crawled corpora, which can involve heterogeneous and unevenly curated data sources. Data ownership structures in this sector are also relatively complex, and access to high-quality domain-specific data may depend on external licensing arrangements. By contrast, cultural-technology activities, such as digital publishing, film and television IP, and online education, operate within more clearly defined copyright regimes and standardized content formats. These institutional conditions contribute to comparatively structured data environments and facilitate commercialization models (e.g., metadata-based API services) in which data quality is more directly linked to revenue realization. The sector also tends to exhibit integrated toolchains spanning content creation, review, distribution, and analytics, supporting more systematic data governance practices.

The sector-specific results for *Data Asset Management Capability, Data Quality Standard Conformity, and Data Asset Benefit Realization Capability* are presented in Figure 2.

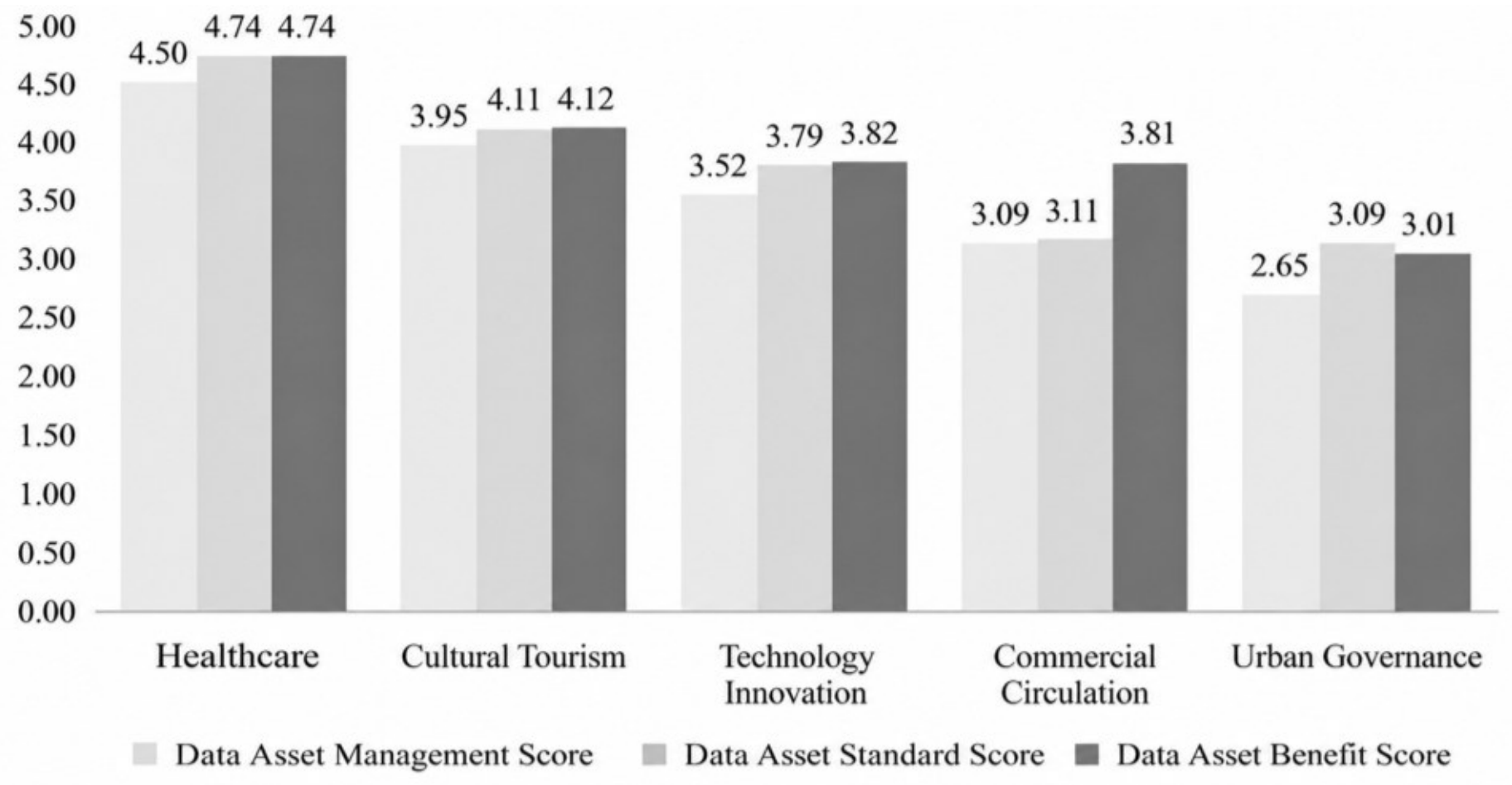

Figure 2. Scores for Data Asset *Management, Standard, and Benefit* Across the Five “Data Elements ×” Sectors

Across the five sectors, the mean scores are 3.54 for *Data Asset Management Capability*, 3.77 for *Data Quality Standard Conformity*, and 3.90 for *Data Asset Benefit Realization Capability*. The within-sector ranges are 1.85, 1.65, and 1.72, respectively. These results indicate moderate overall development across the three dimensions, with greater dispersion within sectors than between sectors. In all sectors, management scores are lower than those for standard and benefit realization. This pattern suggests that firms may emphasize data application and performance outcomes more than long-term systematic governance capability development.

The Healthcare sector continues to record the highest scores across dimensions. The Urban Governance sector reports the lowest overall score, with cross-departmental coordination within the management dimension scoring 2.65. This may reflect challenges related to data responsibility allocation and institutional coordination in administrative contexts. In the Commercial Circulation sector, the effectiveness score (3.81) exceeds both *Management* (3.09) and *Standard* (3.11). This pattern is consistent with earlier findings that emphasize revenue-oriented priorities over systematic governance investment in this sector. Although the Technology Innovation sector is highly data-intensive in R&D activities, its overall score remains moderate. The scores for *Benefit* (3.82) and *Management* (3.52) do not substantially exceed the cross-sector average. This may indicate constraints in translating data resources into sustained innovation performance, potentially related to commercialization pathways and ecosystem dependencies.

# 5. Formation Mechanisms and Configurational Pathways in Data Asset Quality

In complex management and social-science research, a single analytical approach often provides only a partial account of causal relationships. Partial Least Squares Structural Equation Modeling (PLS-SEM) is effective in estimating linear net effects among variables, but is limited in capturing asymmetry and the coexistence of multiple causal paths. Necessary Condition Analysis (NCA) is designed to identify conditions that must be satisfied for an outcome to reach a certain level; however, it focuses on the constraining role of individual conditions and is less suited to uncovering complex combinations of multiple conditions. In contrast, fuzzy-set Qualitative Comparative Analysis (fsQCA) is well suited to identifying equifinal configurations under different combinations of conditions, but does not provide clear estimates of average net effects or relative influence among variables. Recent research increasingly advocates the integration of multiple methods to enhance the robustness and explanatory power of causal analysis [46]. For example, Chen et al. and Sobgo et al. combine PLS-SEM and fsQCA to identify both net-effect relationships and configurational paths, thereby revealing synergistic and substitutive relationships among variables [47,48]. Sukhov et al. further incorporate NCA into this analytical framework to identify necessary but not sufficient conditions alongside net effects and configurational mechanisms [49].

Based on this, this study develops an integrated analytical framework combining PLS-SEM, NCA and fsQCA. First, PLS-SEM is employed to examine the structural relationships among *Management*, *Standard*, and *Benefit*, thereby identifying the direction and relative strength of the effects of different capability dimensions on data asset quality formation. Second, NCA is applied to identify the key capability thresholds required for achieving high levels of data asset quality, both for the overall sample and across five industry categories defined by "data element × industry" contexts. This analysis reveals the constraining role of different capability dimensions and addresses the limitation of structural models in identifying necessary but not sufficient conditions. Finally, fsQCA is used to identify multiple configurational paths leading to high levels of data asset quality from a configurational perspective, thereby capturing heterogeneous mechanisms of capability combinations across different resource endowments and industry contexts. Taken together, this multi-method integration proceeds in a sequential manner—from net-effect estimation, to necessity analysis, to configurational explanation—thereby providing a more comprehensive account of the complex causal mechanisms underlying enterprise data asset quality.

## 5.1 Model Validation and Structural Paths (PLS-SEM)

### 5.1.1 Model Construction and Path Hypotheses

Partial Least Squares Structural Equation Modeling (PLS-SEM) is a variance-based multivariate path modeling technique suitable for complex structural models and relatively small sample contexts [50]. Its flexibility and computational efficiency make it appropriate for models involving multiple latent constructs and moderate sample sizes. No control variables are included to preserve model parsimony and maintain methodological distinction from subsequent configurational analyses. PLS-SEM is primarily employed to identify the main structural relationships among the three dimensions—Management, Standard, and Benefit. At the same time, this study seeks to capture firm heterogeneity through subsequent necessary condition analysis (NCA) and configurational analysis (fsQCA), thereby examining the influence of heterogeneity on outcome formation from multiple analytical perspectives. This approach does not fully substitute for conventional control-variable specifications. Accordingly, the estimated structural relationships should be interpreted as identifying main effects within the given sample context, rather than as fully purified causal estimates that account for all sources of heterogeneity. PLS-SEM is applied to examine the path relationships among three core latent constructs: *Data Asset Management Capability*, *Data Quality Standard Conformity*, and *Data Asset Benefit Realization Capability*. Based on the measurement design, a structural model comprising three second-order reflective constructs is specified. A total of 23 observed indicators are modeled as first-order measures nested under their respective higher-order constructs, forming a second-order reflective PLS path model.

### 5.1.2 Measurement and Structural Model Assessment

The model estimation and path analysis were conducted using SmartPLS 4.0. All indicator loadings exceed the recommended threshold of 0.708 [51], supporting item reliability. $Cronbach's\ \alpha$, composite reliability (CR), and $rho_A$ all exceed 0.90, indicating satisfactory internal consistency reliability.

Convergent validity is assessed using the Average Variance Extracted (AVE). AVE values range from 0.758 to 0.804, exceeding the recommended threshold of 0.50 [51]. These results support convergent validity. Detailed results are presented in Table 6.

Table 6. Measurement Model Results

| Loadings | Data Asset Management Capability (M) | Data Quality Standard Conformity (S) | Data Asset Benefit Realization Capability (B) |
|---|---|---|---|
| Data Strategy (M1) | 0.917 | | |
| Data Governance (M2) | 0.885 | | |
| Data Architecture (M3) | 0.899 | | |
| Data Application Capability (M4) | 0.902 | | |
| Data Security & Privacy Protection (M5) | 0.905 | | |
| Data Quality Management (M6) | 0.932 | | |
| Data Standards & Specifications (M7) | 0.920 | | |
| Data Lifecycle Management (M8) | 0.916 | | |
| Data Integrated Environment (M9) | 0.782 | | |
| Accuracy (S1) | | 0.913 | |
| Consistency (S2) | | 0.881 | |
| Completeness (S3) | | 0.863 | |
| Compliance (S4) | | 0.907 | |
| Timeliness (S5) | | 0.904 | |
| Accessibility (S6) | | 0.857 | |
| Business Impact (B1) | | | 0.807 |
| Decision Support Value (B2) | | | 0.762 |
| Innovation & Opportunity Creation (B3) | | | 0.845 |
| Revenue Generation (B4) | | | 0.842 |
| Cost Reduction (B5) | | | 0.904 |
| Competitive Advantage (B6) | | | 0.948 |
| Risk Management (B7) | | | 0.908 |
| Stakeholder Benefit (B8) | | | 0.931 |
| Cronbach' α | 0.953 | 0.969 | 0.946 |
| rho_A | 0.956 | 0.970 | 0.946 |
| CR | 0.961 | 0.974 | 0.957 |
| AVE | 0.758 | 0.804 | 0.788 |

Discriminant validity is assessed using the heterotrait–monotrait ratio (HTMT). All HTMT values are below the recommended threshold of 0.85, supporting discriminant validity (Table 7). Variance Inflation Factor (VIF) values range from 3.314 to 4.477, remaining below the threshold of 5. With a sample size of 1,500, the mean $R^2$ is 0.843, mean communality is 0.892, and the standardized root mean square residual (SRMR) is 0.063. These indices indicate acceptable explanatory capacity and model fit.

Table 7. Heterotrait-Monotrait Ratio (HTMT) Analysis

| | Data Asset Management Capability (M) | Data Quality Standard Conformity (S) | Data Asset Benefit Realization Capability (B) |
|---|---|---|---|
| Data Asset Management Capability (M) | 1 | 0.848 | 0.840 |
| Data Quality Standard Conformity (S) | | 1 | 0.837 |
| Data Asset Benefit Realization Capability（B） | | | 1 |

### 5.1.3 Core Path Analysis

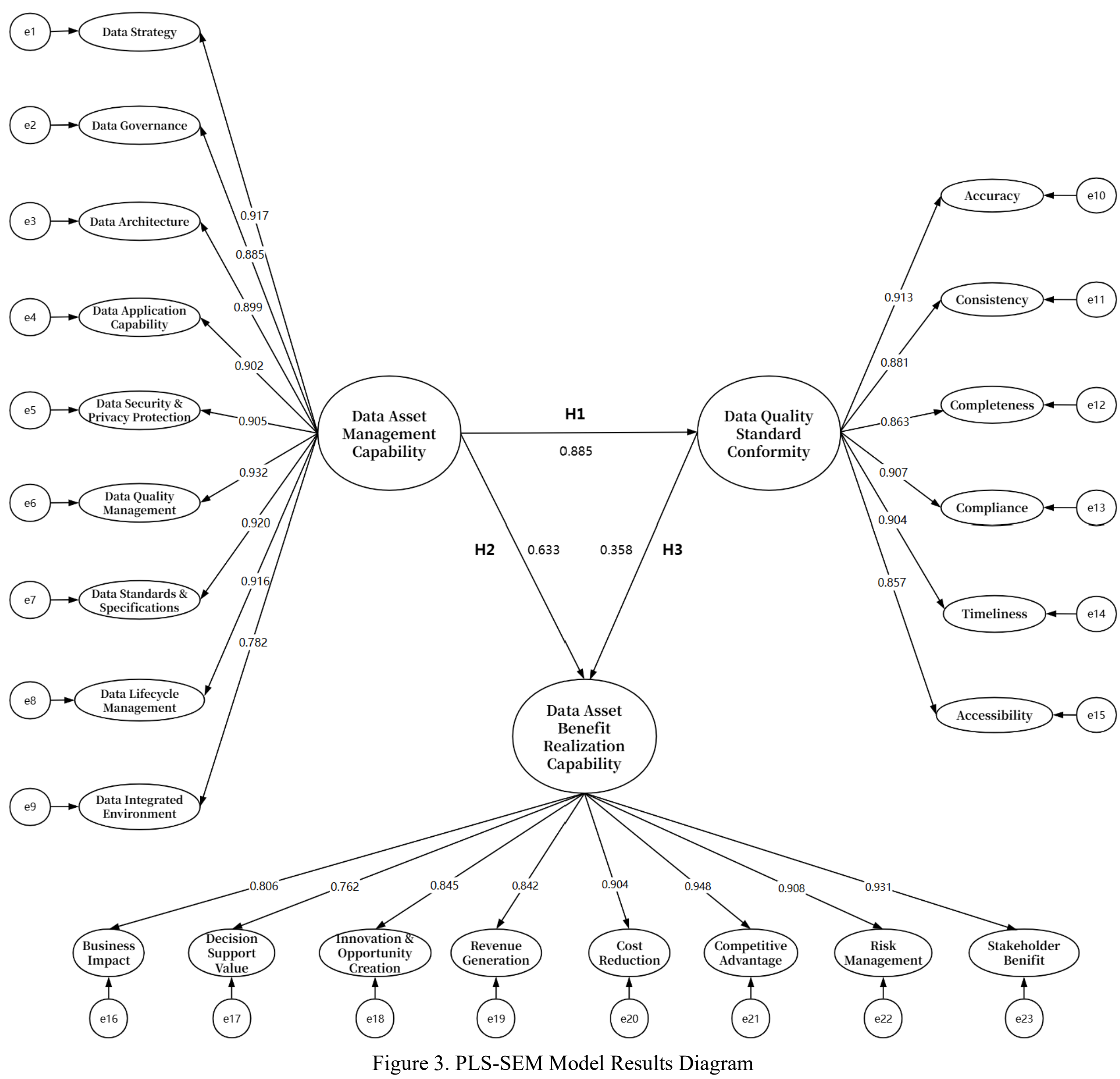

Figure 3. PLS-SEM Model Results Diagram

As shown in Figure 3, all structural paths among *Data Asset Management Capability, Data Quality Standard Conformity, and Data Asset Benefit Realization Capability* are positive and statistically significant ($p < 0.001$), supporting hypotheses H1–H3. Detailed results are reported in Table 8.

Table 8. Path Hypothesis Testing Results

| Path | Direct Effect | p-value | Indirect Effect | Total Effect | $f^2$ |
|---|---|---|---|---|---|
| Data Asset Management Capability → Data Quality Standard Conformity | 0.885 | 0.000*** | - | 0.885 | 0.385 |
| Data Asset Management Capability → Data Asset Benefit Realization Capability | 0.633 | 0.000*** | 0.317 | 0.950 | 0.411 |
| Data Quality Standard Conformity → Data Asset Benefit Realization Capability | 0.358 | 0.000*** | - | 0.358 | 0.301 |

Note: ***, **, and * indicate significance at the 1%, 5%, and 10% levels, respectively. (This notation applies to all subsequent tables.)

Overall, all structural paths are statistically significant ($p < 0.001$), and the effect sizes of Management on Standard ($f^2 = 0.385$) and on Benefit ($f^2 = 0.411$) are relatively high, indicating that the model exhibits strong explanatory power in capturing the relationships among the core variables. The results suggest a chained transmission mechanism of *management foundation → standard enhancement → value realization*, with Standard playing a key mediating role between Management and Benefit. Based on the model outputs and field investigation findings, the main path relationships are interpreted as follows.

First, *Data Asset Management Capability* exerts the strongest direct effect on *Data Quality Standard Conformity*. The path coefficient from Management to Standard is 0.885, indicating that governance capability is the primary driver of the development and effective implementation of data standard systems. As management capability improves, firms tend to simultaneously optimize governance structures, institutionalize quality control processes, and enhance technological system deployment, thereby significantly improving data consistency, normativity, and usability across different systems and business scenarios.

Second, *Data Asset Management Capability* has a significant direct effect on *Data Asset Benefit Realization Capability*. The path coefficient from Management to Benefit is 0.633 (indirect effect = 0.317; total effect = 0.950), suggesting that high-level governance not only determines whether data meet required standards, but also directly affects whether data can be effectively utilized and transformed into performance outcomes such as business impact, decision support, and cost optimization. This finding is consistent with field evidence indicating that "robust governance underpins stable application." Stronger governance capability facilitates the reuse and coordination of data resources within organizations, thereby accelerating the process of value creation.

Finally, *Data Quality Standard Conformity* has a significant positive effect on *Data Asset Benefit Realization Capability* and serves as a transmission mechanism between Management and Benefit. The path coefficient from Standard to Benefit is 0.358, indicating that data quality attributes—such as accuracy, consistency, completeness, timeliness, and accessibility—constitute essential prerequisites for the stable realization of data value. Taken together, the structural relationships indicate that Management not only directly enhances Benefit but also exerts an indirect effect through improving Standard Conformity.

Overall, the improvement of enterprise data asset quality relies on governance capability as the core driver of standard system development, which in turn supports sustained benefit realization through high-quality data supply. These findings provide a structural foundation for the subsequent configurational analysis of multi-path realization mechanisms under different contextual conditions.

**5.1.4 First-Order Loadings Within Second-Order Constructs**

The factor loadings between first-order indicators and their corresponding second-order constructs (Table 9)

indicate the relative contribution of each component to its higher-order construct.

Table 9. Factor Loadings

| Loadings | Data Asset Management Capability (M) | Data Quality Standard Conformity (S) | Data Asset Benefit Realization Capability（B） |
|---|---|---|---|
| Data Strategy (M1) | 0.917 | | |
| Data Governance (M2) | 0.885 | | |
| Data Architecture (M3) | 0.899 | | |
| Data Application Capability (M4) | 0.902 | | |
| Data Security & Privacy Protection (M5) | 0.905 | | |
| Data Quality Management (M6) | 0.932 | | |
| Data Standards & Specifications (M7) | 0.920 | | |
| Data Lifecycle Management (M8) | 0.916 | | |
| Data Integrated Environment (M9) | 0.782 | | |
| Accuracy (S1) | | 0.913 | |
| Consistency (S2) | | 0.881 | |
| Completeness (S3) | | 0.863 | |
| Compliance (S4) | | 0.907 | |
| Timeliness (S5) | | 0.904 | |
| Accessibility (S6) | | 0.857 | |
| Business Impact (B1) | | | 0.807 |
| Decision Support Value (B2) | | | 0.762 |
| Innovation & Opportunity Creation (B3) | | | 0.845 |
| Revenue Generation (B4) | | | 0.842 |
| Cost Reduction (B5) | | | 0.904 |
| Competitive Advantage (B6) | | | 0.948 |
| Risk Management (B7) | | | 0.908 |
| Stakeholder Benefit (B8) | | | 0.931 |

Within the *Data Asset Management Capability* construct, data quality exhibits the highest contribution. Among its indicators, “Data Quality Management” (M6) shows the highest loading (0.932). This indicator captures activities spanning quality objective setting, inspection, analysis, and continuous improvement. The high loading reflects the centrality of quality-related processes within the management construct.

Within the *Data Quality Standard Conformity* construct, data accuracy shows the largest contribution. All observed variables (S1–S6) have loadings above 0.85. Data Accuracy (S1) has the highest loading (0.913). This result indicates that accuracy constitutes a dominant attribute within the standards construct.

Within the *Data Asset Benefit Realization Capability* construct, competitive advantage shows the highest contribution. “Competitive Advantage” (B6) has the highest loading (0.948). The high loading indicates that competitive positioning constitutes a major component of the benefit realization construct.

## 5.2 Necessary Conditions and Capability Thresholds (NCA)

Necessary Condition Analysis (NCA) is a method used to identify “necessary but not sufficient” conditions, aiming to determine whether a given antecedent constitutes a prerequisite for an outcome to reach a certain level [52]. The method emphasizes the minimum conditions required for outcome realization and enables the identification of key capability thresholds associated with high-level outcomes [53,54]. In the preceding PLS-SEM analysis, the structural relationships among *Management*, *Standard*, and *Benefit* have been examined. However, this approach does not allow for assessing which capability dimensions constitute necessary conditions for achieving high levels of data asset quality. To address this limitation, NCA is introduced to identify the constraining role of different capability dimensions in outcome formation.

Specifically, this study primarily employs the CE-FDH technique within the NCA framework. This method is suitable for moderate to large sample sizes and can accommodate potential non-linear relationships with relatively few distributional assumptions, making it appropriate for exploratory research[52,55]. In addition, CR-FDH results are reported for robustness checks. Latent variable scores obtained from the PLS-SEM analysis are used as inputs for the NCA. A ceiling line is drawn above the scatter of observations to identify necessity relationships between predictors and the outcome. The importance of each condition is then evaluated using the NCA effect size (d), and statistical significance is assessed through permutation tests. When $d > 0$, the condition is considered necessary. Specifically, $0 < d < 0.1$ indicates a weak necessary condition, $0.1 \leq d < 0.3$ indicates a moderate level, $0.3 \leq d < 0.5$ indicates a strong level, and $d \geq 0.5$ indicates a very strong necessary condition [52]. On this basis, bottleneck analysis is conducted to identify the minimum levels of each condition required for achieving different levels of data asset quality, both for the overall sample and across five industry categories. This analysis reveals the variation in capability thresholds across dimensions in the formation of high-level outcomes.

#### 5.2.1 Overall Sample Analysis

Table 10. Effect Size Results for the Overall Sample

| | CE-FDH | CR-FDH | p-value |
|---|---|---|---|
| Management | 0.451 | 0.443 | 0.000*** |
| Standard | 0.512 | 0.490 | 0.000*** |
| Benefit | 0.514 | 0.480 | 0.000*** |

The results of the necessary condition analysis for the overall sample are reported in Table 10. Under the CE method, *Benefit* $(d = 0.514)$ and *Standard* $(d = 0.512)$ exceed the threshold of 0.5, indicating very strong necessary conditions. *Management* $(d = 0.451)$ meets the criterion for a strong necessary condition. Together, these three dimensions constitute key necessary conditions for achieving high levels of data asset quality. Permutation tests show that all necessary condition effects are statistically significant $(p < 0.001)$, indicating that the identified necessity relationships are robust.

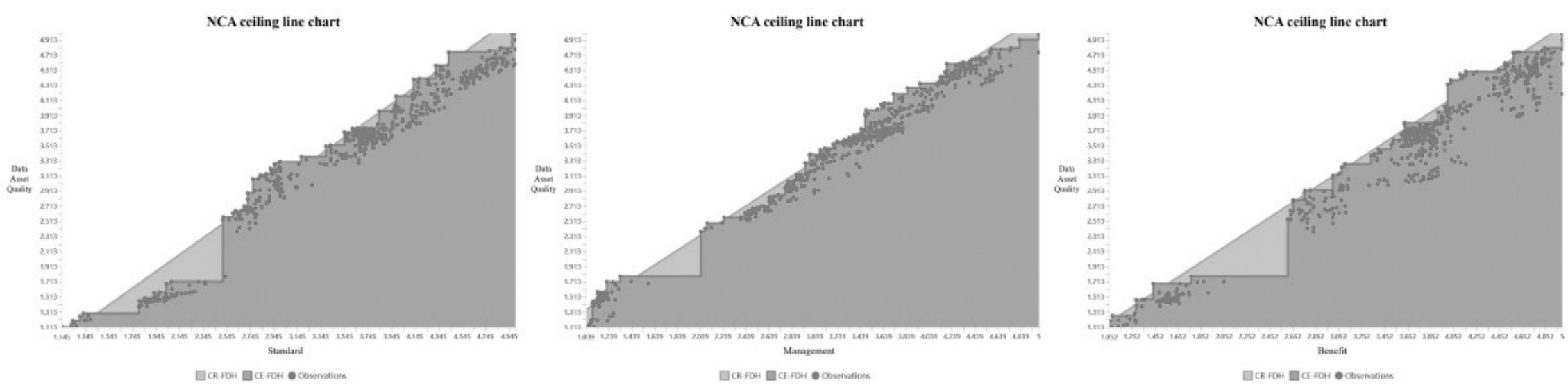

Figures 4–6. Ceiling Line Plots for the Overall Sample

The ceiling line plots in Figures 4–6 show that the upper-left region contains very few data points, indicating that when any capability dimension is at a low level, firms struggle to achieve high levels of data asset quality. This finding highlights the clear necessary condition constraint relationship, which is consistent with the results of the effect size analysis presented in Table 10. It further validates the key constraining role of *Management*, *Standard*, and *Benefit* in the process of data asset quality formation.

Table 11. Bottleneck Table for the Overall Sample

| | Management | Standard | Benefit |
|---|---|---|---|
| 0% | NN | 1.279 | NN |
| 10% | 1.200 | 1.630 | 1.377 |
| 20% | 1.596 | 1.981 | 1.770 |
| 30% | 1.992 | 2.332 | 2.162 |
| 40% | 2.389 | 2.683 | 2.555 |
| 50% | 2.785 | 3.034 | 2.947 |
| 60% | 3.182 | 3.385 | 3.339 |
| 70% | 3.578 | 3.735 | 3.732 |
| 80% | 3.975 | 4.086 | 4.124 |
| 90% | 4.371 | 4.437 | 4.517 |
| 100% | 4.767 | 4.788 | 4.909 |

Table 11 shows that as the target level of data asset quality increases, the minimum required levels of each capability dimension rise accordingly. When data asset quality reaches 50%, the minimum required levels of *Management*, *Standard*, and *Benefit* are 2.785, 3.034, and 2.947, respectively. When the target exceeds 80%, the thresholds for all three dimensions approach or exceed 4.0. Among the three dimensions, *Standard* and *Benefit* exhibit relatively higher thresholds, suggesting that they play a more constraining role in achieving high levels of data asset quality.

**5.2.2 Industry-Specific Analysis under "Data Element × Industry" Contexts**

Existing studies applying Necessary Condition Analysis (NCA) have primarily focused on identifying threshold conditions at the overall sample level, with relatively limited attention to variations in necessity structures across different contexts. To examine such variation, this study further conducts NCA separately for five industry categories—healthcare, urban governance, commercial circulation, technological innovation, and cultural tourism—in order to identify differences in capability constraints under distinct industrial settings. The ceiling line plots for all industry groups exhibit a clear empty region in the upper-left corner, indicating stable necessity relationships among variables, consistent with the effect size results. Accordingly, this study focuses on reporting the effect size estimates and bottleneck analysis results.

5.2.2.1 Healthcare Industry

Table 12. Effect Size Results for the Healthcare Industry

| | CE-FDH | CR-FDH | p-value |
|---|---|---|---|
| Management | 0.451 | 0.367 | 0.000*** |
| Standard | 0.465 | 0.443 | 0.000*** |
| Benefit | 0.438 | 0.389 | 0.000*** |

Table 13. Bottleneck Table for the Healthcare Industry

| | Management | Standard | Benefit |
|---|---|---|---|
| 0% | NN | NN | NN |
| 10% | 4.345 | 4.133 | 4.021 |
| 20% | 4.345 | 4.133 | 4.103 |
| 30% | 4.345 | 4.133 | 4.138 |
| 40% | 4.446 | 4.32 | 4.269 |
| 50% | 4.501 | 4.321 | 4.273 |

(Continuation)

| | Management | Standard | Benefit |
|---|---|---|---|
| 60% | 4.573 | 4.432 | 4.444 |
| 70% | 4.573 | 4.432 | 4.57 |
| 80% | 5.000 | 5.000 | 5.000 |
| 90% | 5.000 | 5.000 | 5.000 |
| 100% | 5.000 | 5.000 | 5.000 |

Table 12 shows that in the healthcare sample, Management $(d = 0.451)$, Standard $(d = 0.465)$, and Benefit $(d = 0.438)$ all exhibit strong necessity and are statistically significant $(p < 0.001)$. The bottleneck analysis in Table 13 indicates that when the target level of data asset quality exceeds 50%, the minimum required levels of all three dimensions approach or exceed 4.3, suggesting relatively high threshold requirements across all capability dimensions in this industry.

5.2.2.2 Urban Governance

Table 14. Effect Size Results for Urban Governance

| | CE-FDH | CR-FDH | p-value |
|---|---|---|---|
| Management | 0.563 | 0.487 | 0.000*** |
| Standard | 0.583 | 0.525 | 0.000*** |
| Benefit | 0.509 | 0.466 | 0.000*** |

Table 15. Bottleneck Table for Urban Governance

| | Management | Standard | Benefit |
|---|---|---|---|
| 0% | NN | NN | NN |
| 10% | 1.435 | 1.928 | 1.141 |
| 20% | 2.608 | 2.513 | 2.391 |
| 30% | 2.608 | 2.513 | 2.391 |
| 40% | 2.703 | 2.693 | 2.526 |
| 50% | 4.000 | 4.105 | 3.48 |
| 60% | 4.000 | 4.105 | 3.48 |
| 70% | 4.000 | 4.105 | 3.48 |
| 80% | 4.000 | 4.255 | 4.499 |
| 90% | 5.000 | 4.969 | 5.000 |
| 100% | 5.000 | 4.969 | 5.000 |

Table 14 indicates that in the urban governance sample, *Management* $(d = 0.563)$, *Standard* $(d = 0.583)$, and *Benefit* $(d = 0.509)$ all reach the level of very strong necessary conditions. The bottleneck analysis in Table 15 further shows that when data asset quality exceeds 50%, the minimum required levels of *Management* and *Standard* already exceed 4.0, suggesting a strong reliance on governance structures and standard systems in this sector.

5.2.2.3 Commercial Circulation

Table 16. Effect Size Results for Commercial Circulation

| | CE-FDH | CR-FDH | p-value |
|---|---|---|---|
| Management | 0.490 | 0.465 | 0.000*** |
| Standard | 0.242 | 0.198 | 0.000*** |
| Benefit | 0.394 | 0.362 | 0.000*** |

Table 17. Bottleneck Table for Commercial Circulation

| | Management | Standard | Benefit |
|---|---|---|---|
| 0% | NN | NN | NN |
| 10% | NN | NN | NN |
| 20% | 3.134 | NN | NN |
| 30% | 3.134 | NN | 2.881 |
| 40% | 3.663 | 2.830 | 2.881 |
| 50% | 3.810 | 2.916 | 2.986 |
| 60% | 3.810 | 2.916 | 3.027 |
| 70% | 3.810 | 2.952 | 3.027 |
| 80% | 4.035 | 2.987 | 3.115 |
| 90% | 4.035 | 3.179 | 3.115 |
| 100% | 4.093 | 3.179 | 3.191 |

Table 16 shows that in the commercial circulation sample, *Management* $(d = 0.490)$ and *Benefit* $(d = 0.394)$ exhibit strong necessity, whereas *Standard* $(d = 0.242)$ shows only moderate necessity. The bottleneck analysis in Table 17 indicates that *Management* has relatively higher threshold requirements for achieving higher levels of data asset quality, suggesting a stronger dependence on organizational and operational capabilities.

5.2.2.4 Technological Innovation

Table 18. Effect Size Results for Technological Innovation

| | CE-FDH | CR-FDH | p-value |
|---|---|---|---|
| Management | 0.562 | 0.488 | 0.000*** |
| Standard | 0.542 | 0.464 | 0.000*** |
| Benefit | 0.503 | 0.456 | 0.000*** |

Table 19. Bottleneck Table for Technological Innovation

| | Management | Standard | Benefit |
|---|---|---|---|
| 0% | NN | NN | NN |
| 10% | NN | 2.715 | 2.479 |
| 20% | 2.752 | 2.724 | 2.654 |
| 30% | 3.000 | 2.841 | 2.957 |
| 40% | 3.103 | 3.158 | 2.957 |
| 50% | 3.509 | 3.388 | 3.202 |
| 60% | 3.626 | 3.694 | 3.456 |
| 70% | 3.997 | 4.000 | 3.586 |
| 80% | 4.701 | 4.315 | 3.808 |
| 90% | 4.701 | 4.411 | 4.150 |
| 100% | 5.000 | 5.000 | 4.194 |

Table 18 indicates that in the technological innovation sample, *Management* $(d = 0.562)$, *Standard* $(d = 0.542)$, and *Benefit* $(d = 0.503)$ all exhibit very strong necessity. The bottleneck analysis in Table 19 shows that as the target level of data asset quality increases, the threshold levels of all dimensions rise accordingly, with *Management* and *Standard* exhibiting particularly strong constraining roles.

5.2.2.5 Cultural Tourism

Table 20. Effect Size Results for Cultural Tourism

| | CE-FDH | CR-FDH | p-value |
|---|---|---|---|
| Management | 0.404 | 0.386 | 0.000*** |
| Standard | 0.382 | 0.37 | 0.000*** |
| Benefit | 0.528 | 0.517 | 0.000*** |

Table 21. Bottleneck Table for Cultural Tourism

| | Management | Standard | Benefit |
|---|---|---|---|
| 0% | NN | NN | NN |
| 10% | 3.599 | 3.544 | 3.287 |
| 20% | 3.599 | 3.544 | 3.287 |
| 30% | 3.599 | 3.544 | 3.287 |
| 40% | 3.627 | 3.544 | 3.446 |
| 50% | 3.916 | 3.842 | 3.924 |
| 60% | 4.273 | 4.141 | 4.104 |
| 70% | 4.611 | 4.736 | 4.178 |
| 80% | 4.765 | 4.836 | 4.300 |
| 90% | 5.000 | 5.000 | 4.829 |
| 100% | 5.000 | 5.000 | 4.995 |

Table 20 shows that in the cultural tourism sample, *Benefit* $(d = 0.528)$ exhibits a very strong necessary condition, while *Management* $(d = 0.404)$ and *Standard* $(d = 0.382)$ show strong necessity. The bottleneck analysis in Table 21 indicates that when data asset quality exceeds 70%, the minimum required levels of all three dimensions approach or exceed 4.5, suggesting that achieving high-quality outcomes relies on the joint enhancement of multiple capability dimensions.

The overall results reveal clear industry heterogeneity in the structure of necessary conditions. In the healthcare sector, relatively balanced requirements are observed across data system construction, standardization, and application. In urban governance, the necessity of *Management* and *Standard* is more pronounced, reflecting the importance of unified standards, data-sharing rules, and institutional arrangements in public-sector contexts. In commercial circulation, *Management* exhibits relatively higher necessity, indicating a stronger reliance on operational capabilities such as supply chain coordination and data integration. In technological innovation, the results suggest a coordinated relationship among data management, technical standards, and innovation outcomes. In cultural tourism, *Benefit* plays a more prominent role, suggesting that value realization relies more heavily on application-oriented data such as user behavior and platform operation data.

### 5.3 Configurational Paths of Data Asset Quality (fsQCA)

The formation of enterprise data asset quality often exhibits complex characteristics, including the coexistence of multiple conditions and asymmetric relationships. To further explore these features, this study employs fuzzy-set Qualitative Comparative Analysis (fsQCA) to identify configurational patterns among different capability dimensions and to examine multiple sufficient condition combinations associated with high levels of data asset quality. This approach provides a more comprehensive understanding of the multi-path mechanisms underlying data asset quality formation and offers insights for firms to select appropriate development strategies under different contextual conditions.

This study uses fsQCA 3.0 to analyze configurations involving *Data Asset Management Capability*, *Standard*, *Benefit*, and overall data asset quality. The analysis follows four main steps: data calibration, necessity testing, sufficiency analysis, and robustness checks. Based on the three-level indicator data from the original survey, weighted scores at both the second- and first-order levels are first calculated using the entropy weight method (EWM), thereby constructing a multi-level evaluation structure. Calibration is conducted using a fuzzy-set approach, transforming continuous variables into membership scores ranging from 0 to 1. The thresholds are set at 0.95 for full membership, 0.5 for the crossover point, and 0.05 for full non-membership [56,57]. A small adjustment (0.001) is added to values at the crossover point to avoid ambiguity. In the configurational analysis, the consistency threshold is set at 0.8, the PRI consistency threshold at 0.75, and the frequency cutoff at 10. Core conditions are defined as those appearing in both the parsimonious and intermediate solutions, while peripheral conditions appear only in the intermediate solution. The consistency of all configurations and the overall solution coverage exceed 0.75, suggesting that the model exhibits satisfactory explanatory power and robustness.

#### 5.3.1 Configurational Sufficiency Analysis of Data Asset Management Capability

5.3.1.1 Calibration and necessity testing

Calibration results for the conditions underlying *Data Asset Management Capability* are reported in Table 22. Single-condition necessity analysis is then conducted to assess whether each condition constitutes a prerequisite for the outcome.

Table 22. Data Calibration for Data Asset Management Capability

| Variable Name | | Calibration | | |
|---|---|---|---|---|
| | | Full Membership | Crossover Point | Full Non-membership |
| Outcome Variable | Management | 0.856 | 0.606 | 0.063 |
| Condition Variables | M1 | 0.899 | 0.589 | 0.021 |
| | M2 | 0.902 | 0.580 | 0.021 |
| | M3 | 0.871 | 0.550 | 0.022 |
| | M4 | 0.909 | 0.632 | 0.093 |
| | M5 | 0.902 | 0.575 | 0.022 |
| | M6 | 0.918 | 0.613 | 0.025 |
| | M7 | 0.926 | 0.611 | 0.023 |
| | M8 | 0.966 | 0.619 | 0.073 |
| | M9 | 0.950 | 0.706 | 0.215 |
| Calibration points: 0.95, 0.5, 0.05 | | | | |

With the necessity-consistency threshold set at 0.90, M1 (Data Strategy), M2 (Data Governance), M3 (Data Architecture), M5 (Data Security & Privacy Protection), M6 (Data Quality Management), M7 (Data Standards), and M8 (Data Lifecycle Management) all exceed 0.90, meeting the criterion for necessity; results are reported in Table 23.

Table 23. Single Condition Necessity Analysis Results for Data Asset Management Capability

| Condition Variables | Necessity Analysis - High Management | | Necessity Analysis - Low Management | |
|---|---|---|---|---|
| | Consistency | Coverage | Consistency | Coverage |
| M1 | 0.914 | 0.909 | 0.559 | 0.469 |
| ~M1 | 0.466 | 0.555 | 0.891 | 0.898 |
| M2 | 0.901 | 0.932 | 0.535 | 0.468 |
| ~M2 | 0.485 | 0.553 | 0.922 | 0.888 |
| M3 | 0.929 | 0.956 | 0.507 | 0.441 |
| ~M3 | 0.457 | 0.523 | 0.950 | 0.919 |
| M4 | 0.866 | 0.968 | 0.465 | 0.439 |
| ~M4 | 0.497 | 0.524 | 0.966 | 0.859 |
| M5 | 0.917 | 0.942 | 0.510 | 0.443 |
| ~M5 | 0.458 | 0.525 | 0.934 | 0.905 |
| M6 | 0.906 | 0.952 | 0.538 | 0.478 |
| ~M6 | 0.504 | 0.564 | 0.946 | 0.894 |
| M7 | 0.943 | 0.969 | 0.519 | 0.451 |
| ~M7 | 0.466 | 0.534 | 0.964 | 0.934 |
| M8 | 0.919 | 0.951 | 0.524 | 0.458 |
| ~M8 | 0.477 | 0.542 | 0.944 | 0.908 |
| M9 | 0.880 | 0.918 | 0.552 | 0.486 |
| ~M9 | 0.507 | 0.573 | 0.906 | 0.865 |

5.3.1.2 Configuration Identification

Based on the results of the necessary condition analysis, and considering antecedent combinations with consistency greater than 0.8, three typical configurations associated with enhanced *Data Asset Management Capability* are identified (Configurations m1-m3, corresponding to the three configurations for high Management). These configurations reflect differentiated governance logics, namely practice-oriented, institution-oriented, and technology-oriented approaches.

**Configuration m1: Practice-Oriented Configuration (Cm1).** Cm1 is characterized by the core presence of Data Application and Data Lifecycle Management, with Data Strategy and Data Governance appearing as peripheral conditions, while Data Integrated Environment is absent as a core condition. This configuration is commonly observed in small- and medium-sized technology firms at the early stage of digital transformation. Its governance characteristics are reflected in a relatively weak emphasis on strategic coordination and a stronger focus on tools and application scenarios. In contexts where formal institutional systems are not yet fully developed, firms tend to establish localized governance mechanisms through iterative “pilot–adjustment” processes. Cm1 illustrates how firms, under resource constraints, can achieve incremental improvements in data asset management through flexible resource allocation.

**Configuration m2: Institution-Oriented Configuration (Cm2).** In this configuration, Data Governance, Data Architecture, Data Application Capability, Data Security, Data Quality Management, Data Standards & Specifications, and Data Integrated Environment all appear as core conditions, forming a coordinated mechanism centered on institutional development. Cm2 is more commonly observed in highly regulated industries such as finance and healthcare. Its governance logic emphasizes the integration of top-level design, process control, and compliance evaluation. By strengthening risk control mechanisms, it simultaneously promotes the standardization and structuring of data asset management.

**Configuration m3: Architecture-Driven Configuration (Cm3).** Cm3 is similarly characterized by the core

presence of multiple technology- and application-related conditions, forming a contrast with the institution-oriented configuration. Its defining feature lies in a technology-driven governance approach, built upon data platforms, automated tools, and flexible system architectures. Through cross-system integration and platform-based governance, firms are able to achieve agile responses and dynamic optimization in data asset management. In this configuration, formal institutional arrangements are relatively less emphasized, with governance focus shifting toward platform capabilities and engineering execution.

The three configurations represent distinct governance logics. Together, they indicate that there is no single optimal path for improving *Data Asset Management Capability*. Instead, firms develop diverse realization mechanisms depending on industry characteristics, organizational stages, and technological foundations.

5.3.1.3 Configurational Analysis Results

For *Data Asset Management Capability*, the raw coverage values of C1, C2, and C3 under High Management are 0.426, 0.710, and 0.713, respectively. Solution consistency equals 0.999, and solution coverage equals 0.751 (Table 24).

Table 24. Configuration Analysis of Data Asset Management Capability

| Condition Variables | High Management Configuration Analysis | | | Low Management Configuration Analysis | | | | |
|---|---|---|---|---|---|---|---|---|
| | Configuration 1 | Configuration 2 | Configuration 3 | Configuration 1 | Configuration 2 | Configuration 3 | Configuration 4 | Configuration 5 |
| M1 | • | | | | ⊗ | ⊗ | ⊗ | • |
| M2 | • | ● | | ⊗ | ⊗ | ⊗ | ⊗ | |
| M3 | | ● | ● | ⊗ | | | ⊗ | ⊗ |
| M4 | ● | ● | ● | ⨂ | ⨂ | ⨂ | ⨂ | ⨂ |
| M5 | | ● | ● | ⊗ | | ⊗ | | ⊗ |
| M6 | | ● | ● | ⊗ | ⊗ | ⊗ | | ⊗ |
| M7 | | ● | ● | ⊗ | ⊗ | ⊗ | ⊗ | ⊗ |
| M8 | ● | | ● | | ⊗ | ⊗ | ⊗ | |
| M9 | ⊗ | ● | ● | ⊗ | ⊗ | | ⊗ | ⊗ |
| Solution Consistency | 0.999 | | | 0.997 | | | | |
| Solution Coverage | 0.751 | | | 0.869 | | | | |

Note: ●and ⨂indicate the presence and absence of core conditions, respectively; •and ⊗indicate the presence and absence of peripheral conditions. (same for all tables)

**5.3.2 Configurational Sufficiency Analysis of Data Quality Standard Conformity**

5.3.2.1 Calibration and necessity testing

The calibration procedure follows the same steps applied to *Data Asset Management Capability*. Calibration results are reported in Table 25.

Table 25. Data Calibration for Data Quality Standard Conformity

| Variable Name | | Calibration | | |
|---|---|---|---|---|
| | | Full Membership | Crossover Point | Full Non-membership |
| Outcome Variable | Standard | 0.934 | 0.684 | 0.247 |
| Condition Variables | S1 | 0.971 | 0.685 | 0.272 |
| | S2 | 0.936 | 0.671 | 0.253 |
| | S3 | 0.934 | 0.683 | 0.310 |
| | S4 | 0.971 | 0.732 | 0.161 |
| | S5 | 0.973 | 0.740 | 0.219 |
| | S6 | 0.973 | 0.606 | 0.170 |
| Calibration points: 0.95, 0.5, 0.05 | | | | |

Single-condition necessity analysis shows that S1 (Accuracy), S2 (Consistency), S3 (Completeness), S4 (Compliance), S5 (Timeliness), and S6 (Accessibility) each exhibit necessity consistency values above 0.90. Results are presented in Table 26.

Table 26. Single Condition Necessity Analysis for Data Quality Standard Conformity

| Condition Variables | Necessity Analysis - High Quality | | Necessity Analysis - Low Quality | |
|---|---|---|---|---|
| | Consistency | Coverage | Consistency | Coverage |
| S1 | 0.949 | 0.955 | 0.457 | 0.436 |
| ~S1 | 0.439 | 0.461 | 0.952 | 0.947 |
| S2 | 0.926 | 0.959 | 0.448 | 0.440 |
| ~S2 | 0.459 | 0.468 | 0.959 | 0.925 |
| S3 | 0.929 | 0.963 | 0.452 | 0.444 |
| ~S3 | 0.464 | 0.472 | 0.963 | 0.927 |
| S4 | 0.930 | 0.960 | 0.464 | 0.453 |
| ~S4 | 0.471 | 0.481 | 0.959 | 0.928 |
| S5 | 0.945 | 0.921 | 0.434 | 0.400 |
| ~S5 | 0.384 | 0.417 | 0.914 | 0.940 |
| S6 | 0.934 | 0.936 | 0.461 | 0.438 |
| ~S6 | 0.439 | 0.463 | 0.933 | 0.930 |

5.3.2.2 Configuration Identification

Based on antecedent combinations with consistency greater than 0.80, four typical configurations associated with high *Data Quality Standard Conformity* are identified (Configurations s1-s4, corresponding to the four configurations for high Standard). These configurations reflect differentiated realization logics of standard capability under varying levels of digital maturity and governance resources.

**Configuration s1 (Cs1): Infrastructure-Oriented Configuration.** Cs1 is characterized by the core presence of Data Consistency, Compliance, and Timeliness, forming a foundational structure of data standards. It is commonly observed in small- and medium-sized enterprises with relatively weak information infrastructure. Given their simplified organizational and system architectures, firms are able to establish a basic governance framework linking standards, processes, and operational efficiency through the introduction of standardized rules and process logic. Cs1 reflects the initial formation of data governance systems.

**Configuration s2 (Cs2): Process-Optimization-Oriented Configuration.** Cs2 is characterized by the core presence of Data Accuracy, Compliance, Timeliness and Accessibility, forming a closed-loop process centered on data collection, processing, and access. It is more commonly observed in high-frequency and high-sensitivity industries such as finance and healthcare. The governance logic emphasizes precision control along the data transmission chain and strong access management capabilities to meet concurrent demands for high responsiveness and accuracy. Cs2 reflects a process-oriented governance pattern.

**Configuration s3 (Cs3): System-Integration-Oriented Configuration.** Cs3 is characterized by the core presence of Data Accuracy, Consistency, Completeness, Compliance and Accessibility, emphasizing systematic control across the entire data lifecycle. It is typically observed in large, multi-business enterprises. In environments involving multiple business lines and cross-platform operations, firms establish multi-dimensional standard systems and interdepartmental coordination mechanisms to mitigate data fragmentation and standard inconsistencies. Cs3 reflects a relatively high level of governance maturity.

**Configuration s4 (Cs4): Agile-Transformation-Oriented Configuration.** Cs4 is characterized by the core presence of Data Accuracy, Consistency, Completeness, Timeliness and Accessibility, highlighting a governance mode driven by the joint effect of foundational data elements and application efficiency. It is commonly observed in internet and manufacturing firms. Through real-time data integration and automated monitoring tools, firms are able to achieve rapid data accumulation and response, thereby strengthening the linkage between governance and operations in dynamic environments.

The four configurations capture distinct patterns of standard capability development. Together, they suggest that improvements in *Data Quality Standard Conformity* follow stage-specific and industry-contingent trajectories, rather than a single uniform path.

5.3.2.3 Configurational Analysis Results

For *Data Quality Standard Conformity*, the raw coverage values of C1, C2, C3, and C4 under High Standard are 0.848, 0.833, 0.806, and 0.806, respectively. Solution consistency equals 0.999, and solution coverage equals 0.909 (see Table 27).

Table 27. Configuration Analysis of Data Quality Standard Conformity

| Condition Variables | High Standard Configuration Analysis | | | | Low Standard Configuration Analysis | | | |
|---|---|---|---|---|---|---|---|---|
| | Configuration 1 | Configuration 2 | Configuration 3 | Configuration 4 | Configuration 1 | Configuration 2 | Configuration 3 | Configuration 4 |
| S1 | | ● | ● | ● | | ⊗ | ⊗ | ⊗ |
| S2 | ● | | ● | ● | ⊗ | | ⨂ | ⨂ |
| S3 | | | ● | ● | ⨂ | ⨂ | | |
| S4 | ● | ● | ● | | ⊗ | ⊗ | | ⊗ |
| S5 | ● | ● | | ● | ⨂ | ⨂ | ⨂ | ⨂ |
| S6 | | ● | ● | ● | | | ⊗ | |
| Raw Coverage | 0.848 | 0.833 | 0.806 | 0.806 | 0.850 | 0.846 | 0.82 | 0.849 |
| Unique Coverage | 0.044 | 0.028 | 0.000 | 0.015 | 0.013 | 0.009 | 0.0099 | 0.0014 |
| Consistency | 0.99 | 0.999 | 0.999 | 0.999 | 0.999 | 0.999 | 0.998 | 0.99 |

（continuation）

| Condition Variables | High Standard Configuration Analysis | | | | Low Standard Configuration Analysis | | | |
|---|---|---|---|---|---|---|---|---|
| | Configuration 1 | Configuration 2 | Configuration 3 | Configuration 4 | Configuration 1 | Configuration 2 | Configuration 3 | Configuration 4 |
| Solution Consistency | 0.997 | | | | 0.997 | | | |
| Solution Coverage | 0.909 | | | | 0.882 | | | |

**5.3.3 Configurational Sufficiency Analysis of Data Asset Benefit Realization Capability**

5.3.3.1 Calibration and necessity testing

The calibration procedure is identical to that described above. Calibration results for *Data Asset Benefit Realization Capability* are reported in Table 28.

Table 28. Data Calibration for Data Asset Benefit Realization Capability

| Variable Name | | Calibration | | |
|---|---|---|---|---|
| | | Full Membership | Crossover Point | Full Non-membership |
| Outcome Variable | Benefit | 0.934 | 0.697 | 0.155 |
| Condition Variables | B1 | 0.970 | 0.733 | 0.208 |
| | B2 | 0.976 | 0.652 | 0.356 |
| | B3 | 0.969 | 0.735 | 0.111 |
| | B4 | 0.936 | 0.699 | 0.115 |
| | B5 | 0.931 | 0.682 | 0.126 |
| | B6 | 0.968 | 0.724 | 0.070 |
| | B7 | 0.972 | 0.697 | 0.027 |
| | B8 | 0.968 | 0.701 | 0.022 |
| Calibration points: 0.95, 0.5, 0.05 | | | | |

Single-condition necessity analysis indicates that B3 (Innovation & Opportunity Creation), B4 (Revenue Generation), B5 (Cost Reduction), B6 (Competitive Advantage), B7 (Risk Management), and B8 (Stakeholder Benefit) each present necessity consistency values above 0.90. Detailed results appear in Table 29.

Table 29. Single Condition Necessity Analysis for Data Asset Benefit Realization Capability

| Condition Variables | Necessity Analysis - High Benefit | | Necessity Analysis - Low Benefit | |
|---|---|---|---|---|
| | Consistency | Coverage | Consistency | Coverage |
| B1 | 0.892 | 0.909 | 0.511 | 0.452 |
| ~B1 | 0.463 | 0.522 | 0.898 | 0.878 |
| B2 | 0.893 | 0.945 | 0.504 | 0.463 |
| ~B2 | 0.493 | 0.53 | 0.940 | 0.88 |
| B3 | 0.936 | 0.904 | 0.568 | 0.476 |
| ~B3 | 0.458 | 0.550 | 0.886 | 0.922 |
| B4 | 0.933 | 0.925 | 0.565 | 0.486 |
| ~B4 | 0.483 | 0.561 | 0.913 | 0.922 |
| B5 | 0.931 | 0.935 | 0.544 | 0.474 |
| ~B5 | 0.478 | 0.547 | 0.926 | 0.92 |
| B6 | 0.940 | 0.949 | 0.557 | 0.488 |
| ~B6 | 0.494 | 0.562 | 0.941 | 0.931 |

（continuation）

| Condition Variables | Necessity Analysis - High Benefit | | Necessity Analysis - Low Benefit | |
|---|---|---|---|---|
| | Consistency | Coverage | Consistency | Coverage |
| B7 | 0.957 | 0.917 | 0.567 | 0.471 |
| ~B7 | 0.448 | 0.544 | 0.901 | 0.948 |
| B8 | 0.954 | 0.953 | 0.581 | 0.503 |
| ~B8 | 0.503 | 0.580 | 0.946 | 0.947 |

5.3.3.2 Configuration Identification

Based on antecedent combinations with consistency greater than 0.80, four typical configurations associated with high *Data Asset Benefit Realization Capability* are identified (Configurations b1-b4, corresponding to the four configurations for high Benefit). These configurations capture distinct strategic orientations and resource integration modes in the process of data value realization, revealing multiple pathways through which firms generate value from data assets.

**Configuration b1 (Cb1): Strategy-Oriented Configuration**. Cb1 is characterized by the core presence of Decision Support Value and Competitive Advantage, with Business Impact and Revenue Generation appearing as peripheral conditions, reflecting a benefit realization logic centered on internal capability development. It is commonly observed in knowledge-intensive firms. The underlying mechanism lies in leveraging data to enhance managerial decision-making, reduce trial-and-error costs, and improve organizational resilience. Even with a relatively limited reliance on direct market-based returns, firms are able to accumulate capabilities and indirectly enhance performance. Cb1 reflects a decision-oriented value realization pattern, in which data contributes to outcomes primarily through internal organizational mechanisms.

**Configuration b2 (Cb2): Co-Creation-Driven Configuration.** Cb2 is characterized by the core presence of Decision Support Value, Competitive Advantage, and Stakeholder Benefit, with Business Impact and Innovation & Opportunity Creation appearing as peripheral conditions, reflecting a multi-actor value co-creation mechanism. Firms activate collaborative relationships within platform ecosystems through strategic coordination and data-driven advantages. For example, e-commerce platforms utilize data feedback to optimize supply-side decisions while strengthening upstream and downstream coordination. In this configuration, data sharing not only supports firm-level objectives but also functions as a key mechanism for ecosystem interaction.

**Configuration b3 (Cb3): Ecosystem-Locking Configuration**. Cb3 is characterized by the core presence of Competitive Advantage and Stakeholder Benefit, with Business Impact, Innovation & Opportunity, and Risk Management appearing as peripheral conditions. Compared with Cb2, Cb3 places greater emphasis on stability within existing ecosystem structures and risk control. For instance, in connected vehicle ecosystems, data exchange between automobile firms and insurance institutions forms a closed-loop value system. Through refined pricing and behavioral feedback, firms achieve coordinated gains while maintaining a balance between compliance and profitability. Cb3 reflects how firms use data to strengthen ecosystem stability and enhance the resilience of value generation.

**Configuration b4 (Cb4): Leveraged-Integration Configuration**. Cb4 is characterized by the core presence of Decision Support Value, Competitive Advantage, and Stakeholder Benefit, complemented by peripheral conditions such as Innovation & Opportunity, Revenue Generation, Cost Reduction, and Risk Management, forming a multi-dimensional value integration structure. It is commonly observed in fintech firms and data platform-based organizations. Through risk modeling, platform interfaces, and service integration, firms are able to extend and replicate data-driven value. Cb4 reflects a composite mechanism of resource integration and capability amplification.

The four configurations represent distinct value realization logics, including decision-driven, ecosystem-oriented, and competition-enhancing pathways. Together, they indicate that firms can achieve improvements in *Data Asset Benefit Realization Capability* through multiple strategic approaches depending on their developmental stage and industry context, rather than relying on a single dominant path.

5.3.3.3 Configurational Analysis Results

For *Data Asset Benefit Realization Capability*, the raw coverage values of C1, C2, C3, and C4 under High Benefit are 0.793, 0.801, 0.823, and 0.768, respectively. Solution consistency exceeds 0.99, and solution coverage equals 0.896 (Table 30).

Table 30. Configuration Analysis of Data Asset Benefit Realization Capability

| Condition Variables | High Benefit Configuration Analysis | | | | Low Benefit Configuration Analysis | | | |
|---|---|---|---|---|---|---|---|---|
| | Configuration 1 | Configuration 2 | Configuration 3 | Configuration 4 | Configuration 1 | Configuration 2 | Configuration 3 | Configuration 4 |
| B1 | • | • | • | | ⊗ | ⊗ | ⊗ | ⊗ |
| B2 | ● | ● | | ● | ⊗ | ⊗ | | ⊗ |
| B3 | | • | • | • | ⊗ | | ⊗ | ⊗ |
| B4 | • | | | • | ⊗ | | ⊗ | ⊗ |
| B5 | | | | • | | ⊗ | ⊗ | |
| B6 | ● | ● | ● | ● | ⊗ | ⊗ | ⊗ | ⊗ |
| B7 | | | • | • | | ⊗ | ⊗ | ⊗ |
| B8 | | ● | ● | ● | | ⊗ | ⊗ | ⊗ |
| Raw Coverage | 0.793 | 0.801 | 0.823 | 0.768 | 0.756 | 0.732 | 0.725 | 0.382 |
| Unique Coverage | 0.023 | 0.0013 | 0.0299 | 0.0145 | 0.0542 | 0.0302 | 0.0238 | 0.0432 |
| Consistency | 0.997 | 0.9979 | 0.9989 | 0.9999 | 0.9984 | 0.9995 | 0.9998 | 0.9988 |
| Solution Consistency | 0.995 | | | | 0.997 | | | |
| Solution Coverage | 0.869 | | | | 0.853 | | | |

**5.3.4 Configurational Sufficiency Analysis of Enterprise Data Asset Overall Quality**

5.3.4.1 Calibration and necessity testing

The calibration procedure remains unchanged. Calibration results for Enterprise Data Asset Overall Quality are reported in Table 31.

Table 31. Data Calibration for Enterprise Data Asset Overall Quality

| Variable Name | | Calibration | | |
|---|---|---|---|---|
| | | Full Membership | Crossover Point | Full Non-membership |
| Outcome Variable | Overall Quality | 0.898 | 0.657 | 0.157 |
| Condition Variables | Management | 0.856 | 0.605 | 0.062 |
| | Standard | 0.933 | 0.683 | 0.247 |
| | Benefit | 0.933 | 0.697 | 0.154 |
| Calibration points: 0.95, 0.5, 0.05 | | | | |

Single-condition necessity analysis shows that *Management*, *Standard*, and *Benefi*t each exceed the 0.90 necessity-consistency threshold. Results are reported in Table 32.

Table 32. Single Condition Necessity Analysis for Enterprise Data Asset Overall Quality

| Condition Variables | Necessity Analysis - High Overall Quality | | Necessity Analysis - Low Overall Quality | |
|---|---|---|---|---|
| | Consistency | Coverage | Consistency | Coverage |
| Management | 0.965 | 0.956 | 0.528 | 0.451 |
| ~Management | 0.446 | 0.523 | 0.949 | 0.959 |
| Standard | 0.933 | 0.976 | 0.474 | 0.428 |
| ~Standard | 0.452 | 0.499 | 0.973 | 0.927 |
| Benefit | 0.947 | 0.95 | 0.553 | 0.478 |
| ~Benefit | 0.480 | 0.555 | 0.942 | 0.939 |

5.3.4.2 Configuration Identification

Two types of configurational patterns are identified in this dimension (Configurations o1–o2, corresponding to the two configurations for high Overall Quality), reflecting two distinct governance logics: internal control-driven and value-feedback-oriented approaches. Antecedent combinations with consistency above 0.80 are retained. Two sufficient configurations associated with high Enterprise Data Asset Overall Quality are identified.

**Configuration o1 (Co1): Internal Control-Driven Configuration**. Co1 is characterized by the peripheral presence of Data Asset Management and Data Asset Standard, with no Benefit-related condition appearing in the configuration. It suggests that firms can improve overall data asset quality by strengthening internal governance systems, even in the absence of prominent benefit-oriented drivers. Co1 is commonly observed in highly regulated sectors, where data systems are built around compliance, stability, and controllability through institutional arrangements and standardized procedures. For example, mechanisms such as access control, log monitoring, and quality auditing are employed to enhance data traceability and clarify accountability. Co1 emphasizes that improvements in data asset quality are achieved on the basis of strengthened governance foundations, in contexts where direct application-driven forces are relatively weak.

**Configuration o2 (Co2): Value-Feedback-Oriented Configuration.** Co2 is characterized by the peripheral presence of Data Asset Standard and Data Asset Benefit, while Data Asset Management does not appear as an explicit prior condition. This does not imply the absence of management as a foundational factor; rather, it suggests that in high-maturity contexts, governance mechanisms may already be embedded within existing institutional arrangements and therefore do not appear as explicit antecedents. Firms associated with this configuration are typically at an advanced stage of data application. They rely on well-established data standards to support value realization and reinvest the resulting benefits into governance tools and platform capabilities, forming a dynamic cycle of standard–value–reinvestment.

5.3.4.3 Configurational Analysis Results

For Enterprise Data Asset Overall Quality, the raw coverage values of Co1 and Co2 under high Overall Quality are 0.903 and 0.901, respectively. Solution consistency equals 0.999 and 0.998, and overall solution coverage equals 0.927 (Table 33).

Table 33. Configuration Analysis of Enterprise Data Asset Overall Quality

| Condition Variables | High Overall Quality Configuration Analysis | | Low Overall Quality Configuration Analysis |
|---|---|---|---|
| | Configuration 1 | Configuration 2 | Configuration 1 |
| Management | ● | | ⊗ |
| Standard | ● | ● | ⊗ |
| Benefit | | ● | |
| Raw Coverage | 0.903 | 0.901 | 0.923 |
| Unique Coverage | 0.027 | 0.023 | 0.923 |
| Consistency | 0.999 | 0.998 | 0.994 |
| Solution Consistency | 0.997 | | 0.994 |
| Solution Coverage | 0.927 | | 0.923 |

# 6. Conclusion and Recommendations

## 6.1 Research Conclusions

This study develops an evaluation framework for enterprise data asset quality by integrating grounded theory, analysis of standard-related texts, and multi-method empirical validation. Under different "Data Element × Industry" contexts, a mixed-method approach (PLS-SEM → NCA → fsQCA) is employed to systematically reveal the underlying mechanisms of data asset quality formation and optimization from three perspectives: structural relationships, necessary condition constraints, and configurational pathways.

At the level of evaluation results, the empirical analysis shows that the overall data asset quality score of sampled enterprises in Hubei Province is 3.82, indicating an intermediate level, yet without a stable high-quality pattern. Comparative results across dimensions reveal that *Data Quality Standard Conformity* (3.81) outperforms *Data Asset Management Capability* (3.51) and *Data Asset Benefit Realization Capability* (3.66). This suggests that firms have made relatively greater progress in establishing standard rules, while there remains room for improvement in systematic governance capacity and the efficiency of value realization. Across most industries, the Management dimension scores lower than both the Standard and Benefit dimensions, reflecting a tendency toward application-oriented over governance-oriented practices. This indicates that firms are more inclined to pursue short-term gains through data utilization, while underinvesting in long-term institutional development.

Regarding sectors differences, significant variation in data asset quality levels is observed. The healthcare industry records the highest score (4.62), reflecting the advantage of coordinated development between standardized systems and governance capabilities under a highly regulated environment. In contrast, the urban governance sector exhibits the lowest score (2.86), indicating that public-sector contexts still face institutional constraints in data coordination and incentive mechanisms. Notably, the cultural tourism industry (4.03) outperforms the technology and innovation sector (3.66) on several dimensions, suggesting that high data dependency does not necessarily correspond to high-quality governance. In the technology and innovation sector, tensions exist between rapid data expansion and the need for refined governance, reflecting challenges in resource allocation. By contrast, the cultural and creative industries benefit from clearer intellectual property delineation and more established standard systems, which facilitate the joint improvement of data quality and value realization.

At the level of structural mechanisms, the results confirm significant positive relationships among *Data Asset Management Capability*, *Data Quality Standard Conformity*, and *Data Asset Benefit Realization Capability*. The PLS-SEM results indicate that the path from Management to Standard exhibits the strongest effect (0.885, $p < 0.001$), suggesting that governance capability serves as a key driver in the development and effective implementation of data standards. Management capability also exerts a direct effect on Benefit Realization Capability (0.633), while additionally generating an indirect effect through enhancing Standard Conformity (indirect effect = 0.317). Meanwhile, the path from Standard to Benefit is also significant (0.358). Taken together, these findings suggest that the improvement of enterprise data asset quality follows a transmission logic of *management foundation → standard enhancement → value realization*. In this process, management capability functions as a foundational prerequisite, standard systems act as a key mediating mechanism, and benefit realization represents the ultimate manifestation of value.

At the level of necessary condition constraints, the NCA results indicate that *Data Asset Management Capability*, *Data Quality Standard Conformity*, and *Data Asset Benefit Realization Capability* all constitute important capability thresholds for achieving high levels of data asset quality. In the full sample, *Data Quality Standard Conformity* and *Data Asset Benefit Realization Capability* exhibit relatively higher levels of necessity, suggesting that firms are unlikely to attain high data asset quality in the absence of well-established standard systems and stable value realization capabilities. Further industry-level analysis reveals substantial variation in the structure of capability thresholds. In highly regulated sectors such as healthcare, the necessity levels of Management, Standard, and Benefit capabilities are relatively balanced, reflecting the need for coordinated development across institutional governance, data standards, and business applications. In governance- and technology-driven sectors, such as urban governance and technology innovation, *Data Asset Management Capability* and Standard systems exhibit stronger necessity, indicating that unified standards and institutionalized governance are critical foundations for ensuring data quality. By contrast, in application-driven industries such as commerce and cultural tourism, *Data Asset Benefit Realization Capability* plays a more constraining role, suggesting that firms in these sectors rely more heavily on operational and platform-based data to achieve value realization. Overall, the formation of enterprise data asset quality depends not only on foundational conditions such as governance capability and standard systems, but is also shaped by differences in industry business models and patterns of data utilization, resulting in heterogeneous capability threshold structures across sectors.

At the configurational level, the fsQCA results further reveal that multiple combinations of conditions can lead to equivalent high-quality outcomes, reflecting pronounced contextual heterogeneity and path diversity. Specifically, within the dimension of *Data Asset Management Capability*, three configurations—Practice-Oriented, Institution-Oriented, and Architecture-Driven—are identified, corresponding to governance logics driven by application scenarios, institutional coordination, and technological dominance, respectively. Within the dimension of *Data Quality Standard Conformity*, four configurations—Infrastructure-Oriented, Process-Optimization-Oriented, System-Integration-Oriented, and Agile-Transformation-Oriented—are identified, reflecting firms' differentiated choices in constructing standard systems under varying levels of digital maturity and industry environments. Within the dimension of *Data Asset Benefit Realization Capability*, four configurations—Strategy-Oriented, Co-Creation-Driven, Ecosystem-Locking, and Leveraged-Integration—are identified, revealing multiple mechanisms of value realization, including enhanced internal decision-making, deepened ecosystem collaboration, and amplified resource integration. At the overall quality level, two configurations—Internal Control-Driven and Value-Feedback-Oriented—further indicate that firms can either achieve steady improvements in data asset quality by strengthening governance foundations or reinforce standards and governance capabilities through value realization. Overall, the formation of enterprise data asset quality is not driven by a single causal pathway but emerges from the multi-path synergy of governance capability, standard systems, and value realization under different stages of development and resource endowments.

From a comparative perspective across methods, the findings of PLS-SEM, NCA, and fsQCA are broadly consistent in their core conclusions, while differing in their explanatory focus. The PLS-SEM results demonstrate significant positive structural relationships among Management, Standard, and Benefit, with the strongest effect observed for the *Management → Standard* path, thereby identifying the primary direction of net effects among the three dimensions. The NCA results further show that *Management*, *Standard*, and *Benefit* all constitute critical capability thresholds for achieving high levels of data asset quality, indicating that these dimensions not only exert average promoting effects but also function as necessary conditions for outcome realization. The fsQCA results, in turn, reveal that high data asset quality can be achieved through multiple equivalent configurations, such as Internal Control-Driven and Value-Feedback-Oriented pathways, depending on firm-specific contexts. Taken together, the three methods complement each other by capturing net effects, necessary conditions, and configurational pathways, respectively, jointly supporting the overarching explanatory framework of *management foundation → standard enhancement → value realization*.

## 6.2 Recommendations

To foster the development of regional data factor markets and strengthen *Data Asset Management Capability*, the following policy and managerial recommendations are proposed.

### 6.2.1 Prioritize the development of governance capability to address the structural imbalance toward application at the expense of governance infrastructure.

Empirical results show that the Management dimension scores relatively low, while the path from *Management* to *Standard* exhibits the strongest effect (0.885), indicating that governance capability serves as the foundational prerequisite for the effective implementation of standards and the realization of value. Accordingly, it is necessary to strengthen the institutional foundation from a regulatory and organizational perspective. This includes refining rules for data asset ownership recognition and balance sheet inclusion, and clarifying operational guidelines for the separation of ownership, control, and usage rights, as well as accounting standards. In addition, firms should improve their data governance organizational structures by exploring the establishment of Chief Data Officer (CDO) roles and cross-departmental coordination mechanisms, thereby strengthening accountability boundaries and process standardization. At the same time, mechanisms for data governance disclosure and evaluation should be established, incorporating governance maturity into credit and regulatory frameworks. A graded management system can also be explored to create sustained external constraints and internal incentives.

### 6.2.2 Strengthen the role of standard systems as a central hub to reinforce the "*Management–Standard–Benefit*" transmission chain.

Given that the self-reinforcing M–S–B loop is empirically validated across all sectors, targeted efforts should be directed toward accelerating the development of universal technical support infrastructure for data governance. First, the deployment of automated governance tools, such as robotic process automation (RPA), Internet of Things (IoT) systems, and data quality monitoring platforms—should be promoted across production, circulation, and service processes. Second, fiscal incentives and benchmark recognition programs may be introduced for enterprises completing full-chain governance transformation, thereby establishing a positive reinforcement mechanism linking transformation investment, policy incentives, and replicable best practices. Third, the construction of a data standards system centered on traceability, verifiability, and operability should be advanced to improve implementation feasibility and cross-industry compatibility.

### 6.2.3 Enhance data asset value realization capacity to improve the release of data asset benefits.

The evaluation results indicate that there remains room for improvement in the Benefit dimension, while the configurational analysis reveals multiple pathways for value realization. Accordingly, firms should embed data applications into core decision-making and business processes, thereby strengthening the role of data in supporting strategic decisions and competitive advantage. Platform-based firms should be encouraged to develop ecosystem-oriented collaboration structures, enabling multi-party value co-creation through data sharing, interface standardization, and cross-entity coordination. In parallel, efforts should be made to explore data asset securitization and market-based transaction mechanisms, and to refine evaluation indicator systems and supporting institutional arrangements, so as to extend the functionalities of data assets toward measurability, tradability, and financialization. At the technological and market levels, dynamic early warning mechanisms can be established by leveraging blockchain-based data provenance systems and AI-driven data quality monitoring platforms. In addition, pilot programs for data trading matchmaking and support through industry funds can be promoted to enhance the circulation capacity of high-value data products, thereby forming a virtuous cycle of governance investment → value realization → reinvestment.

**6.2.4 Implement differentiated, industry-specific strategies to address sectoral divergence.**

Given the observed variation across "Data Element × Industry" contexts, a stratified and differentiated implementation strategy should be adopted. In highly regulated sectors such as healthcare, it is advisable to further consolidate advantages in regulatory compliance and risk control by strengthening end-to-end quality traceability mechanisms. In governance- and technology-driven sectors, such as urban governance and technology innovation, priority should be given to enhancing data governance systems and technical standard capabilities. In particular, the urban governance sector needs to overcome challenges related to cross-departmental coordination and the delineation of responsibilities, thereby improving the integration efficiency of public data platforms. The technology innovation sector, by contrast, should seek to balance data scale expansion with refined governance, with a focus on strengthening data cleaning and quality control in R&D processes. In application-driven industries such as commerce and cultural tourism, emphasis should be placed on improving value realization capabilities by standardizing data formats, enhancing automated data collection, and deepening platform-based data operations, thereby integrating business data with the process of transforming data into economic and strategic assets. At the same time, cross-cutting capability building should be promoted. This includes supporting industry–academia–research collaboration and the development of data governance alliances, refining classification-based operational guidelines, and providing SMEs with SaaS-based governance tools and professional training. Such measures can lower the barriers to digital transformation and enhance the overall regional capacity for data asset governance.

## 6.3 Limitations and Future Research

Despite constructing a relatively systematic evaluation framework for enterprise data asset quality and conducting mechanism identification and configurational analysis, this study still has several limitations that warrant further research.

First, at the level of data collection, the sample is primarily drawn from data element–oriented enterprises in Hubei Province, with data largely based on self-reported assessments. Although this sample reflects typical practices within the region, certain imbalances remain in terms of industry distribution, firm size, and governance maturity. As such, the external validity of the findings has clear boundaries. For example, the PLS-SEM model does not include an extensive set of control variables, and the estimated structural relationships may be influenced by both common method bias and sample heterogeneity. It should also be noted that the transferability of the proposed framework mainly lies in its evaluation logic, structural design, and indicator system, rather than in the direct comparability of effect sizes or path coefficients across different regions, industries, or institutional contexts. Future research could incorporate multi-regional, multi-industry, and multi-source data to further test the applicability and robustness of the framework, and combine web-scraped data, platform data, and survey data for cross-validation, thereby enhancing the objectivity and diversity of data sources.

Second, at the level of model specification, the mixed-method approach (PLS-SEM → NCA → fsQCA) enables the identification of complex causal mechanisms from the perspectives of structural relationships, capability thresholds, and configurational pathways. However, it does not incorporate a temporal dimension, and thus is limited in capturing the dynamic evolution of data asset quality during firms' digital transformation processes. Future research may introduce time-series mixed models or event-based analytical approaches to explore path dependence and stage-specific transitions in the evolution of data asset quality.

Third, at the level of framework variables, although a three-dimensional structure of Management–Standard–Benefit and a multi-level indicator system have been established, latent factors such as data culture, organizational resilience, and innovation mechanisms have not yet been systematically incorporated. Future studies could integrate in-depth firm interviews and organizational behavior theories to enhance the explanatory power of the framework in complex governance contexts.

Finally, at the application level, although this study incorporates both semiotic perspectives and the DCMM framework to enrich the evaluation dimensions, the feedback loop among evaluation, application, and adjustment remains underdeveloped. Future research could explore embedding this framework into actual enterprise governance processes, and, in combination with intelligent evaluation platforms and policy feedback mechanisms, develop a dynamic digital governance support system to promote the continuous application and institutionalization of evaluation outcomes.

## Data Availability Statement

The data that support the findings of this study are not publicly available due to confidentiality agreements with participating enterprises and institutional restrictions. Aggregated data may be available from the corresponding author upon reasonable request.

## Declaration of Interests

The authors declare that they have no known competing financial interests or personal relationships that could have appeared to influence the work reported in this paper.

# Appendix A. Detailed Definitions of Indicators in the Data Asset Quality Assessment Framework

## I. Data Asset Management Capability

### 1.Data Strategy

Refers to the organization's top-level planning and implementation framework for data acquisition, integration, governance, and application. It reflects the alignment between data management objectives and overall corporate strategy, including planning, execution, and evaluation mechanisms.

### 2.Data Governance

Denotes the institutional arrangements that define roles, responsibilities, and processes related to data management. It reflects the degree of organizational formalization and accountability in managing data assets.

### 3.Data Architecture

Represents the technical and structural design of data models, distribution frameworks, integration systems, and metadata management. It reflects the organization's capability to structure, standardize, and scale data systems.

### 4.Data Application Capability

Refers to the organization's ability to transform data resources into analytical insights, decision support, and service outputs. It captures the effectiveness of converting data into operational and strategic value.

### 5.Data Security & Privacy Protection

Denotes the institutional and technical capacity to safeguard data confidentiality, integrity, and availability across the data lifecycle. It reflects the organization's ability to manage compliance and mitigate data-related risks.

### 6.Data Quality Management

Refers to systematic mechanisms for setting, monitoring, and improving data quality standards. It reflects the organization's capability to ensure reliability and usability of data assets.

### 7.Data Standards & Specifications

Denotes the degree of standardization in terminology, master data, reference data, and indicator systems. It reflects semantic consistency and cross-departmental interoperability.

### 8. Data Lifecycle Management

Refers to the capability to manage data throughout its lifecycle, from creation and development to maintenance and retirement. It reflects process control, traceability, and long-term governance continuity.

### 9. Data Integrated Environment

Denotes the organizational context supporting data management, including leadership commitment, resource allocation, institutional support, and cultural embedding of data-driven practices.

## II. Data Quality Standard Conformity

### 1.Accuracy

Refers to the extent to which data correctly represent real-world business conditions. It reflects data validity and credibility for decision-making purposes.

### 2.Consistency

Denotes the degree to which data remain logically and semantically aligned across systems, processes, and time. It reflects the organization's ability to eliminate conflicts and redundancy.

### 3.Completeness

Refers to the extent to which required data elements are fully recorded without critical omissions. It reflects data sufficiency for analytical and operational use.

**4.Compliance**

Denotes the conformity of data processes and formats to internal regulations and industry standards. It reflects institutionalized and rule-based data management practices.

**5.Timeliness**

Refers to the degree to which data is updated and processed in a timely manner. It reflects the ability of data systems to support dynamic and real-time decision-making.

**6.Accessibility**

Denotes the extent to which authorized users can efficiently retrieve and interpret data when needed. It reflects usability, documentation support, and appropriate access control.

## III. Data Asset Benefit Realization Capability

**1.Business Impact**

Refers to the contribution of data assets to core business performance, including growth, efficiency, and market expansion. It reflects the operational influence of data on enterprise outcomes.

**2.Decision Support Value**

Denotes the extent to which data provide informational support for strategic planning, resource allocation, and risk mitigation. It reflects data-driven enhancement of decision quality.

**3.Innovation & Opportunity Creation**

Refers to the role of data in enabling product innovation, business model transformation, and collaborative expansion. It reflects data-driven dynamic capability and strategic renewal.

**4.Revenue Generation**

Refers to the ability of data assets to directly increase revenue through value creation, optimization, and new business development. It reflects the linkage between data utilization and financial returns.

**5.Cost Reduction**

Denotes the capacity to reduce operational and resource costs through data-enabled optimization and automation. It reflects efficiency gains attributable to data management.

**6.Competitive Advantage**

Refers to the strategic role of data assets in strengthening market positioning, differentiation, and long-term competitiveness. It reflects sustainable value creation beyond short-term performance.

**7.Risk Management**

Denotes the use of data to identify, assess, and mitigate operational, compliance, and supply chain risks. It reflects enhanced organizational resilience and uncertainty control.

**8. Stakeholder Benefit**

Refers to the broader value created by data assets for multiple stakeholders, including customers, employees, shareholders, and partners. It reflects spillover effects and shared-value realization beyond direct corporate profit.